\newcommand\ignore[1]{}
\documentclass[12pt,citesort]{article}
\usepackage{amsmath,amssymb,graphicx,url}
\usepackage{color}
\usepackage{axodraw}

\newcommand{\Half}{ {\frac{1}{2} } }

\newcommand\be{\begin{equation}}
\newcommand\ee{\end{equation}}
\newcommand\bea{\begin{eqnarray}}
\newcommand\eea{\end{eqnarray}}
\newcommand{\bdm}{\begin{displaymath}}
\newcommand{\edm}{\end{displaymath}}
\newcommand\nn{ \nonumber\\}

\newcommand{\E}[1]{e^{\textstyle #1}}

\makeatletter
\@addtoreset{equation}{section}
\makeatother

\title{Implications of multi-Regge limits for the Bern-Dixon-Smirnov conjecture}
  
\author{ Richard  C. Brower\footnote{Physics Department,
Boston University, Boston MA 02215},
Horatiu Nastase\footnote{Global Edge Institute, Tokyo Institute of
Technology, Tokyo 152-8550,Japan},  
Howard J. Schnitzer\footnote{Theoretical Physics Group, Martin Fischer School of 
Physics,
Brandeis University, Waltham, MA 02454} \\
and  \\ Chung-I Tan\footnote{Physics Department, Brown University,
Providence, RI 02912}
}

\begin{document}
 
\maketitle

\begin{abstract}
 Planar ${\cal N} =4$ super Yang-Mills $SU(N)$ theory is expected to
  exhibit stringy behavior, 
  anticipated by the 't Hooft genus expansion and the $AdS/CFT$
  correspondence. We examine the Bern-Dixon-Smirnov (BDS) conjecture
  for $n$-gluon amplitudes in the context of single-Regge and multi-Regge
  limits and show  that these amplitudes have the expected Regge form in the Euclidean region. 
  
\end{abstract}

\newpage
\section{Introduction}

The $AdS/CFT$ correspondence, in conjunction with recent work,
has made it possible to study $n$-gluon scattering amplitudes in the
$SU(N)$ planar ${\cal N} =4$ super Yang-Mills (SYM) theory, both at weak and
strong coupling.  In this context Bern, Dixon and Smirnov (BDS) \cite{Bern:2005iz}
(see also \cite{Anastasiou:2003kj}) have
made a conjecture for the color-ordered maximal helicity-violating (MHV)
$n$-gluon amplitudes. Further, Alday and Maldacena (AM) presented a
strong-coupling description of the $n$-gluon amplitude using the $AdS/CFT$
correspondence~\cite{Alday:2007hr}. Schematically their solution is of the form
\be \label{eq:npoint}
{\cal A}_n = A_{tree} e^{\textstyle -\frac{\sqrt{\lambda}}{2 \pi} A_{min}(C_n)}
\ee
where $A_{min}(C_n)$ is the minimal area for a surface in $AdS_5$
spanning the contour $C_n$ made of $n$ light-like segments with the
$i$th  side given by the on-shell gluon momentum $k^\mu_i$. Alday and
Maldacena pointed out that after regularization, the finite part of
(\ref{eq:npoint}) for $n=4$ agrees with the BDS conjecture. Motivated
by this, it was conjectured that there exists a duality between the
gluon amplitudes and light-like Wilson loops also at weak-coupling, and proven at 1 loop in 
\cite{Drummond:2007aua,Brandhuber:2007yx}.
This duality has been verified at two-loops for $n=4$ and $n=5$, and a
conformal Ward identity was derived for the light-like Wilson loops
$W(C_n)$, presumed to be valid to all orders in the 't Hooft coupling, $\lambda= g^2
N$\cite{Drummond:2007cf,Drummond:2007au}. 
The Ward identity fixes the finite part of the Wilson loop for
$n=4$ and $n=5$, up to an additive constant, and agrees with the BDS
conjecture for these amplitudes.  Recently doubt has been cast on the
BDS conjecture for $n \ge 6$. In particular AM~\cite{Alday:2007he} argued that for
a large number of gluons the Wilson loop disagrees with the BDS
conjecture, although their results might still be compatible with a
possible duality between the gluon amplitudes and the Wilson
loops. There has also been explicit consideration of the $n=6$
amplitude, where problems with the BDS conjecture are
encountered. Astefanesei et. al.~\cite{Astefanesei:2007bk} analyze the strong-coupling
prediction of the BDS conjecture for $n=6$ and find discrepancies with
the $AdS/CFT$ prescription of AM. The finite part of the BDS $n=6$
amplitude was also compared to the hexagonal light-like Wilson loop at
two loops \cite{Drummond:2007bm},
with the two expressions differing by a non-trivial function
of the three (dual) conformal invariant variables.

Since large N $SU(N)$ ${\cal N} = 4$ SYM is closely related to a
string theory by means of the $AdS/CFT$ correspondence, it is
reasonable to expect that ${\cal N} =4$ SYM theory exhibits evidence
of stringy behavior. Further this is already anticipated by 't Hooft's
large N expansion of the theory as a genus expansion. In flat-space
string theory, it is well-known that scattering amplitudes exhibit
Regge behavior at high energy with fixed momentum transfer, $t$, e.g., for
2-to-2 scattering,
\be
A(s,t) \sim \beta(t) (s/t)^{\alpha(t)}
\ee
where $\alpha(t)$ is the Regge trajectory function, see Fig. \ref{fig:4regge}.
Indeed the $n=4$ BDS
gluon amplitude can be  recast  \cite{Drummond:2007aua,Naculich:2007ub} so as to
exhibit Regge behavior, with a Regge trajectory for the gluon and
Regge residue given to all orders in perturbation theory in terms of
the cusp anomalous dimension \cite{Korchemsky:1985xj,Ivanov:1985np}.  Given the AM results, 
strong-coupling
limits of the trajectory function and residue are included as well. 

Given the view that ${\cal N} =4$ SYM has string behavior and the Regge behavior of the $n=4$ gluon amplitude, it is suggestive that the gluon scattering amplitudes for $n \ge 5$ might also be expected to  exhibit Regge and multi-Regge behavior.  It is the objective of this paper to examine this issue for $n \ge 5$ BDS gluon amplitudes. We find that the BDS amplitudes have the expected Regge and multi-Regge behavior in various Euclidean limits. A crucial tool in establishing this result is an understanding of the cross-ratios in various Regge limits. This may have significance beyond ${\cal   N}=4$ SYM.  

Further, we relate the solution to the conformal Ward identity
for the lightlike polygon Wilson loop proposed in \cite{Drummond:2007cf,Drummond:2007au} to  the  
Regge behavior of the BDS amplitude, giving support to the strong coupling Wilson loop/gluon 
amplitude duality implied by the Alday-Maldacena proposal. (Note that this duality fails at finite 
temperature
\cite{Ito:2007zy}).

In Sec.~\ref{sec:four} we review the BDS conjecture and the Regge behavior of the
$n=4$ amplitude. In Sec.~3 we describe the Regge and multi-Regge limits of 
general $n>4$
amplitudes. 
In Sec.~\ref{sec:five}, we successfully recast the $n=5$ BDS amplitude to
reveal its Regge behavior. In Sec.~\ref{sec:six} we consider  the Regge and multi-Regge 
limits
of the $n=6$ BDS amplitude. Sec.~\ref{sec:general} extends the 
discussion
to the BDS amplitudes for $n \ge 7$. In Sec. \ref{sec:conformal} we relate the Wilson loop solution to 
the conformal Ward identity to the  Regge behavior of the BDS amplitudes. In Sec.~\ref{sec:concl} we 
summarize our results. 
Appendix A describes details of 
the BDS variables and constraints, while Appendix B contains details omitted from Sec.~
\ref{sec:general}.

\section{The four-gluon amplitude}
\label{sec:four}

The Reggeization of the gluon in non-supersymmetric Yang Mills
theories~\cite{Grisaru:1973vw,Grisaru:1974cf,Fadin:1975cb,Fadin:1996tb,Korchemskaya:1996je}, 
as well as supersymmetric Yang Mills
~\cite{Grisaru:1981ra,Grisaru:1982bi,Schnitzer:2007kh,Schnitzer:2007rn}, 
has a long
history. We review this issue for ${\cal N} =4$ SYM in the context of
the BDS conjecture for the on shell 2-2 gluon scattering amplitude, ${\cal A}_4(k_1
+ k_2 \rightarrow - k_3 - k_4)$, where the  Mandelstam variables are $s =
(k_1 + k_2)^2$, $t = (k_1 + k_4)^2$ and $u=(k_1+k_3)^2$, with
$s+t+u=0$. The $1-2-3-4$ color ordered four point amplitude in the BDS conjecture 
is
\be \label{eq:BDSfour}
{\cal A}_4 = { A}_{tree} { A}^2_{div}(s) { A}^2_{div}(t)
\; \E{\textstyle \frac{f(\lambda)}{8} \log^2(s/t) + \widetilde c(\lambda)}
\ee
where $f(\lambda)$ is proportional to the cusp anomalous dimension and in 
$d=4-2 \epsilon$ the IR divergent contribution is 
\be 
{ A}_{div}(s) = \exp\left[  
-\frac{1}{8 \epsilon^2}  f^{(-2)}\left(\lambda(\frac{ \mu^2}{-s})^\epsilon\right)
-\frac{1}{4 \epsilon}  g^{(-1)}\left(\lambda(\frac{ \mu^2}{-s})^\epsilon\right)
\right]
\ee
where $\lambda = g^2 N$, and $\mu$ is a scale introduced with the IR regulator. Expanding in $
\epsilon$, one obtains $f(\lambda)$ and $g(\lambda)$ 
\cite{Bern:2005iz,Alday:2007hr,Korchemsky:1985xj,
Ivanov:1985np,Collins:1989bt,Magnea:1990zb,Catani:1998bh,Sterman:2002qn}
\bea
f(\lambda)&=& (\lambda \frac{d}{d\lambda})^2 f^{(-2)}(\lambda) =
(\lambda \frac{d}{d\lambda}) f^{(-1)}(\lambda)   \\
g(\lambda)&=& (\lambda \frac{d}{d\lambda}) g^{(-1)}(\lambda)  \nonumber
\eea
where $\widetilde c(\lambda)$  is a a constant. In weak coupling,
\bea
f(\lambda) &=& \frac{\lambda}{2 \pi^2} + O(\lambda^2) \nn
g(\lambda) &=& O(\lambda^2) \\
\widetilde c(\lambda) &=& \frac{\lambda}{16 \pi^2} (\frac{4 \pi^2}{3}) + O(\lambda^2) 
\nonumber
\eea
Although ${\cal A}_4(s,t)$ is manifestly symmetric in $s\leftrightarrow t$, with ${\cal 
A}_{tree}\sim (s/t)$ for $|s/t|>>1$,  a straight forward calculation gives
\be \label{eq:regge4}
{\cal A}_4(s,t) = \beta(t) (s/t)^{\alpha(t)}
\ee
where the gluon trajectory function~\footnote{CIT would like to thank E. M. Levin for 
  earlier collaboration on the relation of this gluon trajectory, Eq. (\ref{eq:alpha}), to 
the corresponding 
 perturbative QCD calculation under  dimensional
  regularization.  Agreement can be achieved by implementing "maximal 
trancendentality".  } 
   is 
\be \label{eq:alpha}
\alpha(t) =  1 + \frac{1}{4 \epsilon} f^{(-1)}(\lambda) - \frac{1}{4} f(\lambda) \log(-t/
\mu^2) +\Half g(\lambda)
\ee
\begin{figure}[th] 
\begin{center}
  \begin{picture}(203,194) (77,-18)
    \SetWidth{0.5}
    \Photon(125,74)(216,74){5.5}{6}
    \SetWidth{1.0}
    \Line(122,79)(122,140)
    \Line(217,81)(217,142)
    \Line(123,7)(123,68)
    \Line(217,5)(217,66)
    \Text(250,69)[lb]{\large{{$\gamma(t)$}}}
    \SetWidth{0.5}
    \Vertex(217,73){6.4}
    \Vertex(123,75){6.4}
    \Text(77,68)[lb]{\large{{$\gamma(t)$}}}
    \Text(216,160)[lb]{\large{{$k_3$}}}
    \Text(119,159)[lb]{\large{{$k_2$}}}
    \Text(168,10)[lb]{\large{{$s$}}}
    \Text(123,-18)[lb]{\large{{$k_1$}}}
    \Text(216,-14)[lb]{\large{{$k_4$}}}
    \Text(169,92)[lb]{\large{{$t$}}}
  \end{picture}
\end{center}
\caption{Regge amplitude for elastic  4-point amplitude defines
and fixes the trajectory function $\alpha(t)$ and the
Reggeon vertex,  $\gamma(t)$.} 
\label{fig:4regge}
\end{figure}
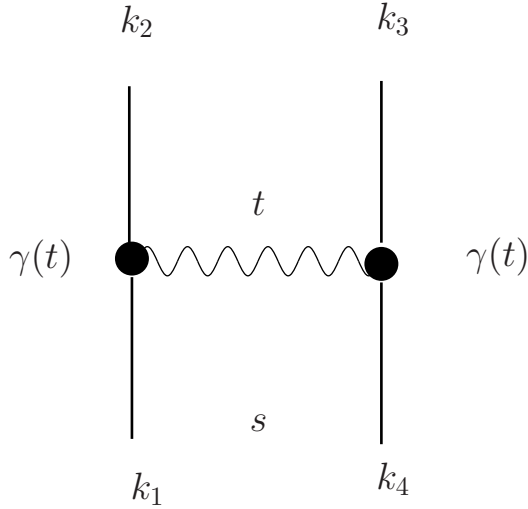

and the Regge residue is
\be \label{eq:beta}
\beta(t) \equiv \gamma^2(t)= \; \mbox{const} \; [{ A}_{div}(t)]^4 \; \E{\widetilde 
c(\lambda)} \; ,
\ee
as in Fig.\ref{fig:4regge}, where  the Sudakov factor
\be
A_{div}(t) = \exp\left\{  -\frac{1}{16 }  f\left(\lambda\right) \log^2(-t/\mu^2) 
 +\left[\frac{1}{8 \epsilon}  f^{(-1)}\left(\lambda\right) + \frac{1}{4 }  g\left(\lambda\right)\right] \log(-t/
\mu^2)  \right\}
 \label{eq:div}
\ee
contains in the exponent a term quadratic in $\log(-t/\mu^2)$ and  the $O(1/\epsilon^2)$ divergent  
constant has been dropped.  One way to see how to go from (\ref{eq:BDSfour}) to (\ref{eq:regge4}) is 
to recognize 
that
\bea \label{eq:s2trel}
{ A}_{div}(s) &=& { A}_{div}(t) \nn
&\times& \exp[  
+\frac{1}{8 \epsilon}  f^{(-1)}(\lambda) \log(s/t) + \frac{1}{4} g(\lambda) \log(s/t) \nn
&-& \frac{1}{16} f(\lambda) \log^2(-s/\mu^2) + \frac{1}{16} f(\lambda) \log^2(-t/
\mu^2)] \; ,
\eea
It is crucial to note  the cancellation of the $\log^2(s/t)$ in Eq.~(\ref{eq:BDSfour})
with the  $\log^2(-s/\mu^2)$ in  ${\cal A}_{div}$ in  leading to
the Regge amplitude (\ref{eq:regge4}).   We also note that Eq. (\ref{eq:regge4}) is first defined in the Euclidean region where $s, t<0$ and then continued to the physical region where $s>0$ and $t<0$, with the phase of the amplitude given by $(s/t )^{\alpha}= e^{-i\pi\alpha} (s /-t)^\alpha$. In our
  subsequent multi-Regge analysis we will not spell out explicitly phases
  resulting from continuation to the physical scattering
  region. The important issue of phase in the
  multi-Regge limit will be consider in subsequent work.

Note that the trajectory (\ref{eq:alpha}) goes as $ - \log(-t/\mu^2)$ rather
than rising linearly with $t$, suggesting stringy behavior, but in the infinite tension
limit. There are no Regge recurrences and no scale for the slope $\alpha'$, 
consistent
with a ${\cal N} =4$ conformal theory with no massive states.~\footnote{HJS thanks 
Lance Dixon for
a discussion of this point.}

\section{Regge limits of $n>4$ amplitudes}
\label{sec:regge}

For $n>4$ amplitudes, the Regge limits are more complicated.  In
particular, one has to distinguish between single-Regge limits, that
are a direct generalization of the Regge limit of the 4-point
amplitude, where a single momentum invariant quantity (the analog of
$s$ for 4-point) goes to infinity, and multi-Regge limits, where
several momentum invariant quantities go to infinity ($s_1,...,s_k$).
Moreover, there are large class of multi-Regge limits.  Here we will
restrict ourselves to the single-Regge and the extreme multi-Regge
example, the so-called ``linear multi-Regge'' limit, since they are
diagrammatically easy to understand and will be used in the general
n-point analysis in Sec.~\ref{sec:general}. For more details on
multi-Regge limits and their realization in flat space open string
theory, see the review \cite{Brower:1974yv}.

\begin{figure}[th]
\begin{center}
  \begin{picture}(332,233) (85,-66)
    \SetWidth{0.5}
    \Photon(197,12)(275,11){3.5}{7}
    \GOval(147,12)(6,49)(0){0.882}
    \GOval(326,11)(6,49)(0){0.882}
    \ArrowLine(104,74)(103,16)
    \ArrowLine(165,74)(164,16)
    \ArrowLine(368,74)(367,16)
    \Text(104,-63)[lb]{\large{{$k_1$}}}
    \Text(225,-16)[lb]{\large{{$t_m$}}}
    \ArrowLine(102,-47)(103,7)
    \ArrowLine(367,-48)(368,6)
    \ArrowLine(280,71)(279,13)
    \ArrowLine(191,74)(190,16)
    \ArrowLine(298,71)(297,13)
    \ArrowLine(322,72)(321,14)
    \ArrowLine(346,73)(345,15)
    \Text(365,-66)[lb]{\large{{$k_n$}}}
    \Text(85,68)[lb]{\large{{$k_2$}}}
    \Text(387,66)[lb]{\large{{$k_{n-1}$}}}
    \ArrowArcn(244,-61.47)(152.47,116.49,63.51)
    \ArrowLine(145,75)(144,17)
    \ArrowLine(121,74)(121,17)
    \ArrowArcn(243.82,3.72)(134.3,147.94,32.56)
    \Text(241,151)[lb]{\large{{$x^2_{i,j}$}}}
    \ArrowArc(220.3,-51.16)(155.38,54.29,125.08)
    \Text(195,109)[lb]{\large{{$x^2_{i,q}$}}}
    \Text(294,107)[lb]{\large{{$x^2_{p,j}$}}}
    \Text(236,73)[lb]{\large{{$x^2_{p,q}$}}}
    \ArrowArc(267.69,-51.11)(155.11,54.39,126.24)
    \Text(233,28)[lb]{\large{{$s_m$}}}
    \DashArrowArc(235.27,4.77)(43.29,29.37,152.14){2}
  \end{picture}
\caption{Cross ratios in the single-Regge limit}
\label{fig:cross_ratio}
\end{center}
\end{figure}
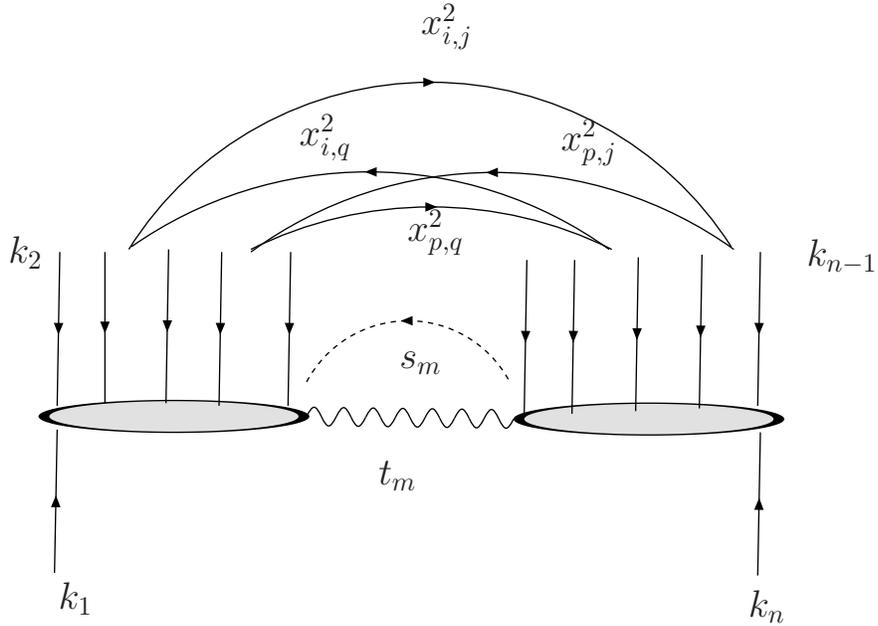

It should be emphasized that the Regge hypothesis, although based on a
long history of experience, is not a proven property in detail,
particularly for a conformal field theory, let alone for ${\cal N} =4$
SUSY. However as one gains confidence in the Regge properties, they do
become an increasingly plausible and powerful non-perturbative
constraint. Indeed this was one of the salient constraints
originally used in the discovery of flat-space string theory.  To be
specific consider the single-Regge limit for a 2-to-(n-2) amplitude as
illustrated in Fig~\ref{fig:cross_ratio}.  The large invariant is
taken to be $s_m = (k_{m+1} + k_{m+2})^2$ with rapidity gap $y \sim
\log(s_m) \rightarrow \infty$.  Without loss of generality, we
consider the group of particles with momenta $k_1,\cdots,k_{m+1}$ to
be ``right movers'' with large postive velocities on the z-axis and
particles with $k_{m+2}, \cdots, k_n$ to be ``left movers'' with
large negative velocities on the z-axis~\footnote{Since we are using an all-incoming convention, 
when ordering the longitudinal components of momenta, we should technically multiply every 
momentum vector $k_j$ by $\pm$, for incoming and outgoing states respectively. To avoid cluttering 
the text, we will not bother to do so, but assume that this will not cause unintended confusion.}.  In 
light-cone coordinates,
the right and left movers have large components $k^+ = (k^0 +
k^3)/\sqrt{2}$ and $k^- = (k^0 - k^3)/\sqrt{2}$ respectively so the
Regge limit can be identified with a matrix element of a Lorentz boost,
\be 
{\cal A}_n(s) \simeq \langle  \bar k_1,\cdots, \bar k_{m+1} |\;  \exp[  yM_{+-}] \; |  \bar k_{m+2}, \cdots
\bar k_n\rangle \; 
\ee 
The matrix element is taken between states boosted to a frame with (near) 
zero z-momenta so that the relative boost back to the original frame is given
by the rapidity difference, $y= y_R - y_L \sim \log(s_m)$.  A detailed analysis~
\cite{Brower:2006ea,Brower:2007qh,Brower:2007xg} leads to the
consequence that the singularities in the angular-momentum $J$-plane are
determined by the spectrum of the boost operator, $M_{+-}$, through the
Mellin-Laplace transform,
\be {\cal A}_n(j) \simeq
\int^\infty_{\mu^2} ds \; s^{- j-1} \; {\cal A}(s)\quad  \sim \quad \langle \bar k_1,\cdots,
\bar k_{m+1} |\frac{1}{j-M_{+-}} | \bar k_{m+2}, \cdots, \bar
k_n\rangle  \; .
\ee
If the largest eigenvalue eigenvalue for $M_{+-}$ is discrete, there is a leading simple pole in the   $J$-plane and a dominant pure power $s_m^{\alpha(t_m)}$ for the amplitude at high energies. A fundamental   hypothesis of Regge theory states that all amplitudes with the same quantum numbers must share exactly the   same $J$-plane poles and the residues factorize \footnote{There is an obvious analogy between the spectral   analysis of the Hamiltonian and the boost oprator using the correspondence $(E,t) \leftrightarrow (-J,y)$.}.  
 
Remarkably the exact 4-point BDS amplitude has been found to corresponds   to a single $J$-plane pole plus integer space  daughters~\cite{Drummond:2007aua,Naculich:2007ub}. This property of an exact   meromorphic $J$-plane also holds for the BDS 5-point function. The similarity between the BDS $J$-plane  and flat space string theory is striking.  The classic 4-point Veneziano   amplitude of flat space string theory is also meromorphic in the   $J$-plane with integer spaced daughters. Since   flat space string theory is integrable, it is  known that there is the absense of Regge cuts in the planar limit, which   raises the question of whether planar ${\cal N}=4$ SYM  might also exhibit pure $J$-plane meromorphy with simple poles and no Regge cuts.

\begin{figure}[th]
\begin{center}
\includegraphics[width = 0.4\textwidth]{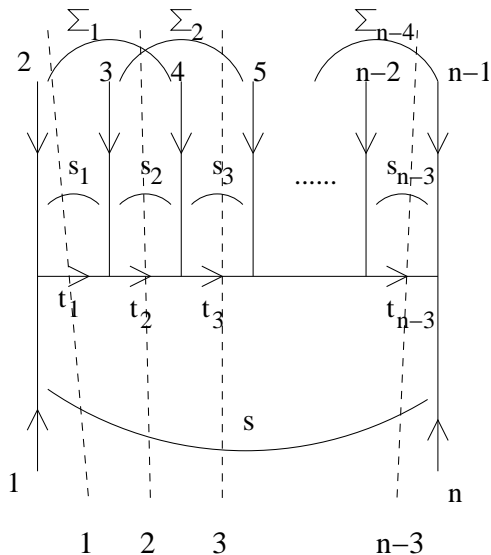}
\end{center}
\caption{n-point amplitude expressed as a tree diagram of effective particle 
("Reggeon") exchange
in order to emphasize the parameterization of the linear multi-Regge limit.}
\label{fig:limits}
\end{figure}

The focus of this article is restricted to the examination of the leading  Regge and multi-Regge behavior for the conjetured BDS amplitudes for   $n>4$ in the Euclidean region. The BDS ansatz for a general $n$-point is conveniently expressed in terms of an over complete set of cyclic ``Mandelstam'' invariants $t_i^{[r]}\equiv (k_i+...+k_{i   +r-1})^2$ for all unitarity cuts of the n-point function with the external legs arranged in the cyclic order of the large N single color trace.  There are $n(n-3)/2$ such distinct cyclic invariants, but due to $n$ mass-shell conditions and 10 Lorentz symmetries, this reduces the $4n$ momentum components, $k^\mu_i$ to $3n - 10$ independent Lorentz variables.  Hence there are constraints for $n\geq 6$. Fortunately in the Regge limit the constraints take a simple form. (see Appendix~\ref{app:BDSvar} for more details on these constraints.)

First let us consider an independent set of invariants most appropriate for the 2-to-(n-2) amplitude in the linear multi-Regge limit. Here one views the $n$-point amplitudes as tree level interactions involving effective particle ("Reggeon") exchange, as in Fig. \ref{fig:limits}. Associated with this configuration, it is natural to define $n-3$ two body energy invariants, $s_i$, and $n-3$ momentum transfers, $t_i$.
\be
s_i=t_{i+1}^{[2]}\; , \quad i=1,...,n-3\;,\quad {\rm and} \quad t_r=t_1^{[r+1]}\;, \; \quad
r=1,...,n-3\; .
\ee
Clearly $\{s_i\}$ and $\{t_r\}$ may be regarded as generalization of $s$ and
$t$ for the 4-point amplitude. In the linear multi-Regge limit, all
$\{s_i\}$ goes to infinity with $\{t_r\}$ fixed.  For the remaining
$n-4$ independent variables, we {\it provisionally} choose the three body
energies,
\be
\Sigma_i=t_{i+1}^{[3]},\;\quad i=1,...,n-4 \;.
\ee
An alternative choice, which is more convenient for discussing various Regge limits, is to scale 
$\Sigma_i$ by defining ratio variables
\be 
\kappa_i \equiv \frac{\Sigma_i}{s_i s_{i+1}} ,\; \quad i=1,...,n-4
\ee 
An n-point amplitude can be considered as a function of this set
of $3n-10$ independent variables, ${\cal A}_n(s_1, \cdots, s_{n-3},
t_1,\cdots, t_{n-3}, \kappa_1,\cdots, \kappa_{n-4})\;$. All other BDS
invariants can be expressed in terms of this set (see
Appendix~\ref{app:BDSvar}.)

{\bf Single Regge Limit:} In this limit there is a single large
rapidity gap, which for the general n-point function we take
to be given by $\log(s_k) \sim \log(s) \rightarrow \infty$, which cuts
the diagram into two halves of right and left movers as described before
and illustrated in Figs.~\ref{fig:limits} and \ref{fig:cross_ratio}. This implies a
Reggeon propagator carrying an invariant mass squared $t_k$ dual to the cut, 
\be
{\cal A}_n \sim s_k^{\alpha(t_k)}
\ee 
A nice pictorial way to
represent the single Regge limit it is to draw a dotted line cutting through the
corresponding ``Reggeon exchange" line, in our example dotted line $k$
cutting through $t_k$ and separating particles $1$ through $k+1$ from the rest, Fig.~\ref{fig:limits}. 
Then any of the BDS invariants, $[i,i+r]$,  that
contains momenta on both sides of the dotted line, (expressed either 
as $t_i^{[r]}$ or as $t_{i+r}^{[n-r]}$, i.e., $[i,i+r]\;({\rm mod} \;n)=t_i^{[r]}=t_{i+r}^{[n-r]}$), will go to
infinity. Thus in this example, from the set of independent variables
$\{s_i,t_j,\Sigma_i\}$, only $s_k$ and $\Sigma_{k-1}, \Sigma_k$ go to infinity (if $k\neq 1$), and
all the others stay fixed.
 Furthermore, since $s_k\sim \Sigma_{k-1}\sim \Sigma_k$,  it follows that $\kappa_k$ is fixed. That is, 
this single-Regge  limit is defined by
$s_k\rightarrow \infty$, with 
$ s_1,s_2\cdots, s_{n-3}, \; (\neq   s_k) , t_1,t_2,\cdots, t_{n-3}$, and $\kappa_1,\kappa_2,\cdots, 
\kappa_{n-4}$ fixed.

As the number of external lines is increased
this leads to a sequence of single-Regge limits,
\bea\label{eq:single}
{\cal A}_4&\simeq& [\gamma(t)]^2 \left(\frac{s}{t}\right)^{\alpha(t)} 
\label{eq:singleregge4}\\
{\cal A}_5 &\simeq& \gamma(t_1) \left(\frac{s_1}
{t_1}\right)^{\alpha(t_1)}G^{[3]}_1(t_1,s_2,\kappa_1,t_2)\label{eq:singleregge5} \\
{\cal A}_6 &\simeq& \gamma(t_1)\left(\frac{s_1}
{t_1}\right)^{\alpha(t_1)}G^{[4]}_1(t_1,\kappa_1,s_2, t_2, \kappa_2, 
s_3,t_3)\label{eq:singleregge6B} \\
{\cal A}_6 &\simeq&G^{[3]}_1(t_1,s_1,\kappa_1,t_2) \left(\frac{s_2}
{t_2}\right)^{\alpha(t_2)}G^{[3]}_1(t_2, s_3, \kappa_2,t_3) \label{eq:singleregge6A}\\
{\cal A}_7 &\simeq & \cdots\cdots \nonumber
\eea
all of which {\bf must} share the same Regge trajectory function,
$\alpha(t)$ and the ``residues'' for different amplitudes factorize
into a single sequence of Reggeon k-particle vertex functions:
$\gamma(t)$, $G_1^{[k]}(t_i, s_i, \kappa)$. This places a strong
recursive consistency condition on the BDS construction.  In essence
this condition reflects the existence of a well defined spectral
decomposition for the boost operator $M_{+-}$ analogous to the
spectral condition for a Hamiltonian.  In particular once $\gamma(t)$
is determined from ${\cal A}_4$ and $G^{[3]}_1$ from ${\cal A}_5$, the
symmetric Regge (\ref{eq:singleregge6A}) limit of ${\cal A}_6$ is
entirely fixed. We show that the BDS conjecture satisfies these
constraints.

As we will see in our subsequent analysis the crucial simplification
of the BDS amplitude in the Regge limits relates to the limit of conformal
cross-ratios~\footnote{Note that the $\kappa$ variables are closely
  related to a cross ratio,
\be
\kappa_{i-2} \; x^2_{i,i+1} =   \frac{x^2_{i,i+1} \Sigma_{i-2}}{s_{i-2} s_{i-1}}
= \frac{x^2_{i,i+1} x^2_{i-1,i+2}}{x^2_{i-1,i+1} x^2_{i,i+2}} \; ,\nonumber
\ee
except that for our present application to ${\cal N}=4$ SYM
amplitudes, the zero mass on-shell conditions, $x^2_{i,i+1} = k^2_i = 0$, must 
be replace by an IR regulator $x^2_{i,i+1} \rightarrow \mu^2$ to
get a finite result.
}. 
The way this works is as follows. The n-point function has momenta
$k_i$ with one energy-momentum constraint, $k_1 + k_2 + \cdots + k_n =
0$. This constraint is satisfied by introducing $n$ variables $x_i$ on
the dual vertices of the dual polygon with $k_i$ assigned to the
edges such that $k_i = x_i - x_{i+1}$ and the variables
\be
x_{i,j} = x_i - x_j = k_i + \cdots + k_{j-1} 
\ee
so that the BDS variables are redefined as unique differences,
\be
t^{[r]}_i = t^{[n-r]}_{i + r }=  x^2_{i,i+r} \equiv (x_i - x_{i+r})^2
\ee
In computation, the notation $[i,j]\equiv x^2_{i,j}$ often proves to be
convenient. All indices are treated cyclically modular
$n$. The BDS amplitudes make special use of conformally invariant
cross ratios,
\be
u(i,j;a,b) = \frac{x^2_{i,j} x^2_{a,b}}{ x^2_{i,b} x^2_{a,j}} = \frac{[i,j] [a,b]}{[i,b] [a,j]} \; .
\ee

As an example in single-Regge limits consider a cross ratio where all
4  invariant factors connect the right movers ($k^+_R \rightarrow \infty$ for $R =
2,\cdots, k+1$) with the left movers ($k^-_L \rightarrow \infty$ for
$L = k+2, \cdots, n-1$) as depicted in Fig.~\ref{fig:cross_ratio}.
The $s_k \rightarrow \infty$ limit implies that all such cross ratios,
\be
u(i,j,p,q)= \frac{x_{i,j}^2x_{p,q}^2}{x_{i,q}^2x_{p,j}^2} 
= \frac{(k_i+...+k_{j-1})^2(k_p+...+k_{q-1})^2}
{(k_i+...+k_{q-1})^2(k_p+...+k_{j-1})^2} \rightarrow
1  + O(1/s_k)
\ee
approach 1.  This follows immediately from the 
fact that scalar products ($p_R q_L\simeq
p^+_R q^-_L) $ between right and left movers become large,
\be
u(i,j,p,q)\simeq \frac{( P^+ Q^-) (p^+ q^-)}{(P^+ q^-) (p^+ Q^-)}    = 1 \; ,
\ee
where we have collected the partial sums into $P^{\mu}=\sum_{R=i}^{k+1} k_R^{\mu}$,
$p^{\mu}=\sum_{R=p}^{k+1}k_R^{\mu}$ and $Q^{\mu}=\sum_{L=k+2}^{j-1}
k_L^{\mu}$, $q^{\mu}=\sum_{L=k+2}^{q-1} k_L^{\mu}$. This is a crucial
kinematic feature leading to Regge behavior for the BDS ansatz. Other
specific examples of the simplification of the constraints in the Regge
limit are given in the following text as we need them and summarized in
Appendix A.

{\bf Linear Multi-Regge Limit:}  
Interpreting the Regge limit in terms of large rapidity separation allows
a systematic generalization to multi-Regge limits
~\cite{Chew:1968fe,Tan:1971mg}. 
The basic idea is to consider the external 
momenta
ordered by rapidities and to separate them in groups with infinite
rapidity differences between them.

The {\em ``linear multi-Regge"} limit (also known as {\em ``multi-peripheral limit"}) 
is defined by taking several infinite
rapidity gaps, corresponding to several of the dotted lines in
Fig. \ref{fig:limits}. The maximal case is when all $n-3$ lines correspond
to infinite rapidity gaps.  Then all of the $s_i$ go to infinity
independently, i.e., the limit is defined by $s_1, s_2,\cdots, s_{n-3}\rightarrow \infty $
holding $t_1,t_2,\cdots, t_{n-3}, \kappa_1, \kappa_2,\cdots, \kappa_{n-4}$ fixed. 
In this limit, since $s$ is linear in each  $s_i$, it can be shown that
\be
s\simeq  \; b_0 s_1s_2\cdots s_{n-3}\;. \label{s-multiregge}
\ee
where $b_0=\Pi_{i=1}^{n-4} \kappa_i$. The postulated behavior for this ``maximal multi-Regge limit'' 
is
\bea
{\cal A}_5&\simeq&  \gamma(t_1) \left(\frac{s_1}{t_1}\right)^{\alpha(t_1)}G_2(t_1, 
\kappa_{1},t_2)  \left(\frac{s_2}{t_2}\right)^{\alpha(t_2)} 
\gamma(t_2)\label{eq:multiregge5} \\
{\cal A}_6 &\simeq& \gamma(t_1) \left(\frac{s_1}{t_1}\right)^{\alpha(t_1)}G_2(t_1, 
\kappa_{1},t_2)  \left(\frac{s_2}{t_2}\right)^{\alpha(t_2)} G_2(t_2, \kappa_{2}, 
t_3) \left(\frac{s_3}{t_3}\right)^{\alpha(t_3)}\gamma(t_3)\nonumber\\
\label{eq:multiregge6-2}\\
& &\cdots\cdots \nonumber\\
{\cal A}_n&\simeq& \gamma(t_1) \left(\frac{s_1}{t_1}\right)^{\alpha(t_1)}G_2(t_1, 
\kappa_{1},t_2)  \cdots\cdots\left(\frac{s_{n-3}}{t_{n-3}}\right)^{\alpha(t_{n-3})}
\gamma(t_{n-3})\label{eq:multireggegeneral}
\eea
In this limit, kinematic simplifications can be achieved since, for $2\leq i<j\leq n$,
\be
[i,j]=x_{i,j}^2= t_i^{[j-i]}=(\sum_{r=i}^{j-1} k_r)^2 \simeq   2k^+_i\cdot k^-_{j-1}+0(1) \;.
\label{eq:strongordering2}
\ee
Moreover, $\kappa_i$ now has  a simple physical interpretation,
\be
\kappa_{i-2} \simeq (2k^+_{i}k^-_{i})^{-1} ={k_{i,\perp}^2} ^{-1}
\ee
so that the sub-energy invariant $s_i$ is proportional to the ratio $k^+_{i+1}/k^+_{i+2}$,
\be
s_i\simeq 2 k^+_{i+1} k^-_{i+2}  \simeq {k_{i+2,\perp}^2}  (k_{i+1}^+/k_{i+2}^+)>>1\;.
\ee
For more details on these kinematic constraints see Appendix~\ref{app:BDSvar}.

\section{The BDS five-gluon amplitude}
\label{sec:five}

The BDS conjectured $n=5$ amplitude for the  on shell gluon amplitude with
$k_1 + k_2 + k_3 + k_4 + k_5  =0$ and $t^{[r]}_i = (k_i + \cdots + k_{i+r-1})^2$
is given by
\be \label{eq:BDS5}
{\cal A} = { A}_{tree} \prod^5_{i=1} A_{div}(t^{[2]}_i) \E{{\cal F}_5(0)}
\ee
where
\bea
{\cal F}_5(0) &=& \frac{f(\lambda)}{8} (L_5 +\frac{15}{2} \zeta_2) \; , \quad
\eea
and 
\be \label{eq:L5}
L_{5}  = - \Half \sum_{i=1}^5\log(\frac{-  t^{[2]}_i}{- t^{[2]}_{i+3}}) 
                   \log(\frac{- t^{[2]}_{i+1}}{- t^{[2]}_{i+2}})
\ee

\begin{figure}[th] 
\begin{center}
  \begin{picture}(380,131) (23,-43)
    \SetWidth{1.0}
    \Line(305,26)(305,60)
    \SetWidth{0.5}
    \Vertex(305,21){5}
    \SetWidth{1.0}
    \Line(360,25)(360,59)
    \SetWidth{0.5}
    \Vertex(361,21){5}
    \Photon(304,22)(256,20){2.5}{6}
    \SetWidth{1.0}
    \Line(361,-19)(361,15)
    \SetWidth{0.5}
    \Photon(307,22)(359,22){2.5}{6}
    \Text(252,-31)[lb]{\large{{$k_1$}}}
    \Text(361,-29)[lb]{\large{{$k_5$}}}
    \Text(302,72)[lb]{\large{{$k_3$}}} %
    \Text(361,70)[lb]{\large{{$k_4$}}}
    \Text(333,62)[lb]{\large{{$s_2$}}}
    \Text(373,15)[lb]{\large{{$\gamma(t_2)$}}}
    \SetWidth{1.0}
    \Line(252,-15)(252,19)
    \Line(252,59)(252,24)
    \Text(246,68)[lb]{\large{{$k_2$}}}
    \Text(275,61)[lb]{\large{{$s_1$}}}
    \Text(219,11)[lb]{\large{{$\gamma(t_1)$}}}
    \Text(300,-5)[lb]{\large{{$G_2$}}}
    \Text(274,4)[lb]{\large{{$t_1$}}}
    \Text(328,4)[lb]{\large{{$t_2$}}}
    \SetWidth{0.5}
    \Vertex(252,22){5}
    \Text(293,-39)[lb]{\large{{$s = \Sigma_1$}}}
    \SetWidth{1.0}
    \Line(107,18)(107,52) %
    \Line(162,21)(162,55)
    \SetWidth{0.5}
    \Photon(106,18)(58,16){2.5}{6}
    \SetWidth{1.0}
    \Line(163,-23)(163,11)
    \Text(54,-35)[lb]{\large{{$k_1$}}}
    \Text(163,-32)[lb]{\large{{$k_5$}}}
    \Text(104,68)[lb]{\large{{$k_3$}}} %
    \Text(163,67)[lb]{\large{{$k_4$}}}
    \Text(135,58)[lb]{\large{{$s_2$}}}
    \Line(54,-19)(54,15)
    \Line(54,55)(54,20)
    \Text(48,64)[lb]{\large{{$k_2$}}}
    \Text(77,57)[lb]{\large{{$s_1$}}}
    \Text(20,7)[lb]{\large{{$\gamma(t_1)$}}}
    \Text(120,-12)[lb]{\large{{$G_1^{[3]}$}}}
    \Text(76,2)[lb]{\large{{$t_1$}}}
    \SetWidth{0.5}
    \Vertex(54,18){5}
    \Text(95,-43)[lb]{\large{{$s = \Sigma_1$}}}
    \GOval(136,16)(6,29)(0){0.882}
  \end{picture}
\end{center}
\caption{Regge limits for 5-point amplitude. On the left, the single Regge limit
factorizes defining a new single Regge 3-particle vertex, $G^{[3]}
(t_1,\kappa_{12},s_2,t_2)$ and on the right, the  double Regge limit
defines a new two-Reggeon vertex, $G_2(t_1,\kappa_1,t_2)$.}
\label{fig:regge5both}
\end{figure}
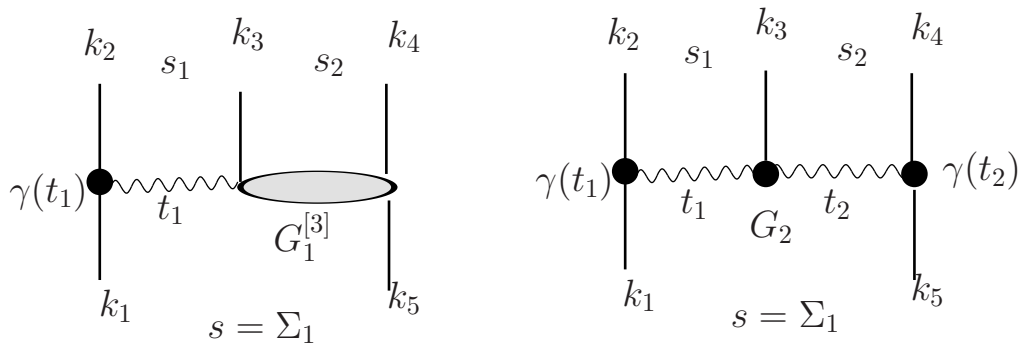

Let us choose a specific kinematical configuration for definiteness,
appropriate to $1+5 \rightarrow \bar 2 + \bar 3 + \bar 4$, as in Fig~
\ref{fig:regge5both}.

The variables  correspond to the BDS variables as follows,
\be
t^{[2]}_1 = t_1  ; \; \;t^{[2]}_2 = s_1 ; \; \;t^{[2]}_3 = s_2 ; \; \;t^{[2]}_4 = t_2 ; \; \;t^{[2]}_5 
= \Sigma_1=s
\ee
As discussed in sec. \ref{sec:regge}, we will use, instead of $\Sigma_1$, an alternative ratio variable
\be
\kappa_1 =  \frac{\Sigma_1}{s_1 s_2}=\frac{s}{s_1 s_2}
\ee
Thus using relations (\ref{eq:s2trel}) in  (\ref{eq:BDS5}) it is a straightforward  
algebraic exercise
to show that the color ordered  5-point BDS amplitude is
\bea \label{eq:5regge}
{\cal A}_5 &=& \mbox{const} \; A^2_{div}(t_1) \;  A_{div}(-\mu^4 \kappa_1) A^2_{div}
(t_2)
(-s_1/\mu^2)^{\alpha(t_1)} (-s_2/\mu^2)^{\alpha(t_2)}(-\mu^4\kappa_1)^{3/2} \nn
&& (-t_1/\mu^2)^{\Half\alpha(\mu^4\kappa_1)-\alpha(-\mu^2)}(-t_2/\mu^2)^{\Half
\alpha(\mu^4\kappa_1)-\alpha(-\mu^2)}
\exp[f(\lambda) \log^2(t_1t_2/\mu^4)/16]\nonumber\\
\eea
where the Regge trajectories appearing in (\ref{eq:5regge}) are the same as the 
gluon trajectory of
(\ref{eq:alpha}.  It should be emphasized that (\ref{eq:5regge}) is an exact 
consequence of Eqs.~(\ref{eq:BDS5}) to (\ref{eq:L5}).

The single Regge limit corresponds to $s_1\rightarrow \infty, s\rightarrow \infty$, 
holding 
$t_1,t_2,s_2,\kappa$ fixed and gives 
\be
{\cal A}_5\sim   (s_1/t_1)^{\alpha(t_1)} \left(\gamma(t_1)G_1^{[3]}(t_1,t_2, s_2, 
\kappa_1)\right)
\ee
as expected.  From (\ref{eq:5regge}) and (\ref{eq:beta}), explicit expression for $G_1^{[3]}(t_1,t_2, s_2, 
\kappa_1)$ can readily be extracted.

According to the general discussion, the double Regge (``multi-Regge") limit
appropriate to (\ref{eq:5regge}) is  taken first with
\be
s_1 \rightarrow - \infty ; \;\; s_2 \rightarrow -\infty ; \;\; s \rightarrow-  \infty 
\ee
while holding $t_1<0 $, $t_2<0$ and $\kappa_1<0$ fixed. The expected form of the
amplitude is given by
\be
{\cal A}_5 = \mbox{const} \; (s_1/t_1)^{\alpha(t_1)} G(t_1,t_2,\kappa_1) (s_2/
t_2)^{\alpha(t_2)} 
\label{a5}
\ee
where we have renamed $G=\gamma(t_1)G_2\gamma(t_2)$.
 It is gratifying to note that (\ref{eq:5regge})
has this factorized form, from which $G_2(t_1,t_2,\kappa_1)$ can also be obtained directly. The physical region is reached by analytically continuing $s_1\rightarrow e^{-i\pi } s_1$, $s_2\rightarrow e^{-i\pi} s_2$ and $\kappa_1\rightarrow e^{i\pi}\kappa_1$.  
The associated analyticity question in $\kappa_1$ is subtle for a conformal theory, and  a careful re-examination  of  the Steinmann rule might  be required~\cite{Brower:1974yv,Detar:1972nd,Cahill:1973px,Tan:1972kr}.  In this paper, we shall focus mainly on Regge behavior in the Euclidean limit.  Again Regge behaviour in (\ref{eq:5regge}) is achieved due 
to the cancellations of the $\log^2(-s_1/\mu^2)$ and $\log^2(-s_2/\mu^2)$ terms in
${\cal F}_5(0)$ with analogous terms in $A_{div}(s_1)$, $A_{div}(s_2)$, and 
$A_{div}(s= \kappa_1 s_1s_2)$ . 

It is worth pointing out that one could naively have expected additional ``Regge-like" limits, e.g.,  (a) 
$s_1$ and $s_2$ becoming large independently, with $t_1$, $t_2$ and  $s$ fixed, 
 or (b) $s_1$, $s_2$, and $s$ becoming large independently, with $t_1$ and $t_2$ fixed. It is easily 
to check that Eq. (\ref{eq:5regge}) would not lead to Regge-like power behavior for these limits. As 
we explain in Sec. \ref{sec:regge}, there is a systematic approach in defining various Regge limits. 
Each limit can be associated with a ``tree-graph". A complex-angular momentum, $J$, can be 
defined relative to  each internal propagator,  and an associated Regge limit defined, with a 
corresponding Regge contribution. Neither  case  (a) nor case (b) listed above is a legitimate Regge 
limit.  These properties can be illustrated explicitly by making use of the Koba-Nielson representation 
for an ordered flat-space open-string 5-point amplitude~\cite{Detar:1972nd}.

Therefore the BDS amplitude has the double Regge form
(\ref{a5}) as expected from a stringy behavior of the 5-gluon
planar ${\cal N}=4$ SYM amplitude.  This supports the conclusion,
reviewed in the Introduction, that the BDS conjecture is in fact valid
for $n=5$.

\section{The BDS six-gluon amplitude}
\label{sec:six}

\begin{figure}[bthp]
\begin{center}\includegraphics{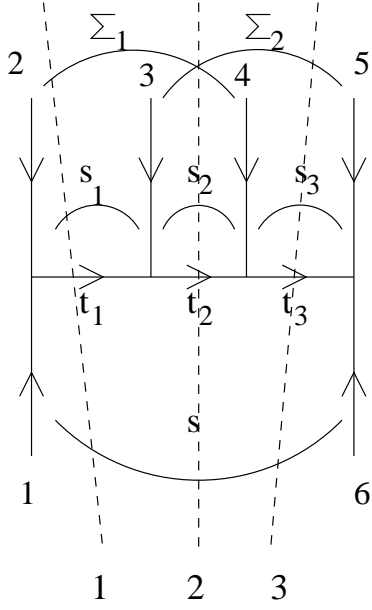}\end{center}
\caption{Linear Regge limits for 6-point gluon amplitude}
\label{fig:6lin}
\end{figure}

We now consider the 6-gluon ordered amplitude. The simplest limit to consider is the single-Regge 
limit defined in section 3,
with variables defined in Fig. \ref{fig:6lin} (from the general case in section 
\ref{sec:regge} and Appendix A).
As we can see from the figure, there are two types of inequivalent single Regge 
limits one can take: type-I, where 
the Regge line is taken to be dotted line 1 or 3, and type-II,
where the Regge line is taken to 
be dotted line 2, with an inelastic vertex on each side of the cut.

The BDS 6-point function is 
\be
{\cal A}_6 = A_{tree} \prod^6_{i=1} A_{div}(t^{[2]}_i) \; \E{{\cal F}_6(0)}
\ee
where
\be
{\cal F}_6(0) = \frac{f(\lambda)}{8} \left( Q_{6} +  D_{6} + L_{6} +  \frac{18}{2} \zeta_2\right)
\ee
with
\bea
Q_{6} &=& - \sum^6_{i=1}  \ln\left( \frac{t^{[2]}_i}{t^{[3]}_{i}}\right)
\ln\left( \frac{t^{[2]}_{i+1}}{t^{[3]}_{i}}\right)\label{eq:Q6}\\
D_{6} &=& - \Half \sum^6_{i=1}  {\rm Li}_2\left(1 - \frac{t^{[2]}_i t^{[4]}_{i-1}}{t^{[3]}_i t^{[3]}_{i-1}}\right) 
\label{eq:d6}\\
L_{6} &=& - \frac{1}{4}  \sum^6_{i=1} \ln\left( \frac{t^{[3]}_i}{t^{[3]}_{i+4}}\right)
\ln\left( \frac{t^{[3]}_{i+1}}{t^{[3]}_{i+3}}\right)=\frac{1}{2}\sum_{i=1}^{3}
\ln^2\left(\frac{-t^{[m]}_{i}}{ -t^{[m]}_{i+1}}\right)
\label{eq:l6}
\eea

For convenience, the BDS variables are now denoted as  $s_i,t_i,\Sigma_i,s$, 
\bea
&& t_1^{[2]}=t_1;\;\; t_2^{[2]}=s_1;\;\; t_3^{[2]}=s_2;\;\; t_4^{[2]}=s_3;\;\;
t_5^{[2]}=t_3;\;\; t_6^{[2]}=s\nonumber\\
&& t_1^{[3]}=t_2;\;\; t_2^{[3]}=\Sigma_1;\;\; t_3^{[3]}=\Sigma_2
\eea
We note that there are 9 BDS variables but one, $s$, will be  considered as a dependent variable~
\footnote{See sec. \ref{sec:regge} and Appendix A for details.}.

For  $n= 6$,  ${\cal F}_6(0)$  involves two new types of terms, $Q_{6}$,  (\ref{eq:Q6}), and $D_{6}$, 
(\ref{eq:d6}). The latter  leads to the presence of dilogarithm function, 
Li${}_2(1-u_i)$,  where for $n=6$  there are three combinations of ``cross ratios", 
\be
u_1= \frac{t_1^{[2]} t_6^{[4]}}{t_1^{[3]}t_6^{[3]}}=\frac{t_1s_3}{t_2\Sigma_2}\; ,\;\quad 
u_2= \frac{t_2^{[2]} t_{1}^{[4]}}{t_2^{[3]}t_1^{[3]}}=\frac{t_3s_1}
{t_2\Sigma_1}\; , \;\quad 
 u_3= \frac{t_3^{[2]} t_2^{[4]}}{t_3^{[3]}t_2^{[3]}}=\frac{s_2s}{\Sigma_1\Sigma_2}\;.
  \label{crossr}
\ee
Although these dilogarithms make the $n=6$ BDS amplitude more involved, we demonstrate below 
that these  terms  do not contribute to the  leading Regge behavior for all the limits we will consider 
here.

Consider first the 
type-I single-Regge limit where
\bea
&&s_1\rightarrow - \infty  \nonumber\\
&&t_1,t_2,t_3,s_2,s_3, \kappa_1, \kappa_2<0 \quad {\rm fixed} 
\label{eq:typeI}
\eea
The dependent variable $s$ also become large, with $s/s_1>0 $  and  fixed.
Note that  all three cross ratios can be expressed as  
\be
u_1= \frac{t_1}{t_2s_2\kappa_2}\; ,\;\quad 
u_2=\frac{t_3}{t_2s_2\kappa_1}\; , \;\quad 
 u_3=  \frac{s }{\kappa_1\kappa_2s_1 s_2s_3}\;.
\ee
and they remain bounded and fixed in this limit. Furthermore, in the Euclidean limit,  dilogarithms $Li_2(1-u_i)$ in (\ref{eq:d6}) will be evaluated on the principal sheet, thus bounded. With $Li_2(0)=0$ and $Li_2(1)=\pi^2/6$, they do not lead to terms which grow with $\log s_1$, thus they 
have no effect on the Regge behavior of the amplitude~\footnote{When continued through the branch cut  of $Li_2(z)$ above  $z=1$, the dilogarithm  becomes  singular at $z=0$, $Li_2(z) \sim  \log \; z $. This does not concern us here but can become important for considering Regge behavior in the physical region.}.
\begin{figure}[th]
\begin{center}
  \begin{picture}(404,98) (-4,-37)
    \SetWidth{0.5}
    \Text(61,43)[lb]{\large{{$p_3$}}}
    \Text(24,44)[lb]{\large{{$p_2$}}}
    \Text(105,43)[lb]{\large{{$p_4$}}}
    \Text(147,45)[lb]{\large{{$p_5$}}}
    \SetWidth{1.0}
    \Line(25,-19)(25,5)
    \Line(63,11)(63,35)
    \Line(144,-24)(144,0)
    \Line(108,10)(108,34)
    \Text(41,33)[lb]{\large{{$s_1$}}}
    \Text(84,34)[lb]{\large{{$s_2$}}}
    \Line(144,9)(144,33)
    \Text(24,-34)[lb]{\large{{$p_1$}}}
    \Text(144,-35)[lb]{\large{{$p_6$}}}
    \SetWidth{0.5}
    \Photon(23,7)(63,8){2}{7}
    \GOval(104,7)(5,42)(0){0.882}
    \Text(38,-11)[lb]{\large{{$t_1$}}}
    \Text(99,-14)[lb]{\large{{$G_1^{[4]}$}}}
    \SetWidth{1.0}
    \Line(25,34)(25,9)
    \Text(284,41)[lb]{\large{{$p_3$}}}
    \Text(247,42)[lb]{\large{{$p_2$}}}
    \Text(328,41)[lb]{\large{{$p_4$}}}
    \Text(370,43)[lb]{\large{{$p_5$}}}
    \Line(248,-21)(248,3)
    \Line(286,9)(286,33)
    \Line(248,32)(248,7)
    \Line(367,-26)(367,-2)
    \Text(304,-10)[lb]{\large{{$t_2$}}}
    \Line(331,8)(331,32)
    \Text(264,31)[lb]{\large{{$s_1$}}}
    \Text(307,32)[lb]{\large{{$s_2$}}}
    \Line(367,7)(367,31)
    \Text(247,-36)[lb]{\large{{$p_1$}}}
    \Text(367,-37)[lb]{\large{{$p_6$}}}
    \SetWidth{0.5}
    \Photon(287,4)(327,5){2}{7}
    \GOval(268,6)(5,20)(0){0.882}
    \GOval(349,3)(5,20)(0){0.882}
    \Text(336,-15)[lb]{\large{{$G_1^{[3]}$}}}
    \Text(262,-16)[lb]{\large{{$G_1^{[3]}$}}}
    \Text(217,4)[lb]{\large{{$\gamma(t_1)$}}}
    \Text(122,35)[lb]{\large{{$s_3$}}}
    \Text(345,33)[lb]{\large{{$s_3$}}}
    \Text(-4,7)[lb]{\large{{$\gamma(t_1)$}}}
  \end{picture}
\caption{Type-I and type-II single-Regge limits for 6-point amplitude.  On the right, type-II single-Regge limit 
with vertices $G_1^{[3]}(t_1,\kappa_1,s_1,t_2)$ and $G_1^{[3]}
(t_3,\kappa_2,s_3,t_2)$, which are dtermined by factorization from the 5-point amplitude.}
\label{fig:multiregge6}
\end{center}
\end{figure}
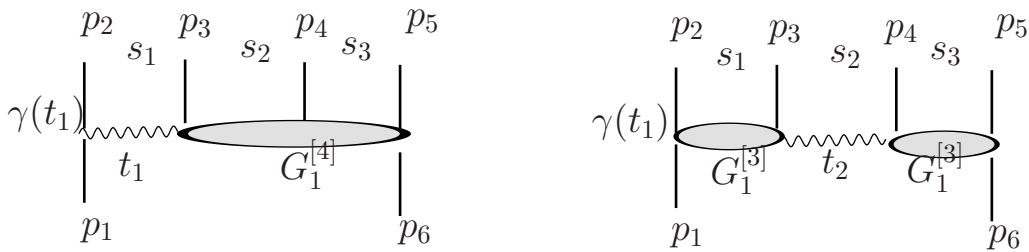

Turning next to   $\log s_1$ terms in $A_{div}$, $Q_{6}$, and $L_{6}$. One can show, by a straight 
forward calculation,  all terms quadratic in $log s_1$ cancel, and the BDS amplitude becomes
\be
\log (A_6/A_{6,tree})\simeq  \log \left(\frac{s_1}{t_1}\right)\left\{ -\frac{f(\lambda)}{4}
\log \left[\frac{-t_1}{\mu^2}\right]+\frac{g(\lambda)}{2}+\frac{f^{-1}(\lambda)}{4\epsilon}\right\}+O(1)
\label{eq:type2bds}
\ee
That is, it has precisely  the desired  Regge behaviour, Eq. (\ref{eq:singleregge6B}),
\be
A_6\sim  \left(\frac{s_1}{t_1}\right)^{\alpha(t_1)} 
\ee
with the same Regge trajectory obtained previously from the $n=4$ BDS amplitude. The $O(1)$ term 
in (\ref{eq:type2bds}) leads to  a new  coupling, $G_1^{[4]}$, Fig. \ref{fig:multiregge6}. To avoid cluttering, we will not exhibit here explicit expression for $G_1^{[4]}$ here.

We consider next   the type-II single-Regge limit, namely
\bea
&&s_2\rightarrow - \infty  \nonumber\\
&&t_1,t_2,t_3,s_1,s_3,\kappa_1,\kappa_2 <0 \quad {\rm fixed}
\label{eq:type2}
\eea
The dependent variable $s$ also become large, now with $s/s_2$  fixed.  In this limit, from 
(\ref{crossr})
and (\ref{uconstra}), we get
\be
u_1\rightarrow 0 \; ,\;\quad 
u_2 \rightarrow 0\;\,\quad 
 u_3\rightarrow 1 \;\;.\label{eq:crossratiosII}
\ee
so that the dilogarithm terms remain bounded and, again, they do not contribute to this Regge limit~
\footnote{With  $u_1\sim u_2\sim 0$, and $Li_2(1-x)-\pi^2/6\sim x\log x$ for $x$ small, these lead to 
$\log s_2/s_2$, thus corresponding to subdominant  Regge  contributions.}.

After canceling quadratic tems in $\log s_2$, we obtain
\bea
&&\log (A_6/A_{6,tree})\simeq  \log \left(\frac{s_2}{t_2}\right)\left\{ -\frac{f(\lambda)}
{8}\left[2
\log \left(\frac{-t_2}{\mu^2}\right)-\log u_3
\right]  
+\frac{g(\lambda)}{2}+\frac{f^{-1}(\lambda)}{4\epsilon}\right\}+O(1)\;.\nonumber\\
\eea
Since $u_3\rightarrow 1$, we have obtained  the expected Regge behavior.  
 We have verified that the 6-point function has  the anticipated  factorized form of Eq. 
(\ref{eq:singleregge6A}), depicted in  Fig. \ref{fig:multiregge6},
\be
{\cal A}_6 \simeq G^{[3]}_1(t_1,s_1,\kappa_1,t_2) \left(\frac{s_2}
{t_2}\right)^{\alpha(t_2)}G^{[3]}_1(t_2, s_3, \kappa_2,t_3)\;. \nonumber
\ee

\begin{figure}
\begin{center}
  \begin{picture}(194,96) (27,-34)
    \SetWidth{1.0}
    \Line(98,12)(98,36)
    \Text(77,34)[lb]{\large{{$s_1$}}}
    \Text(119,35)[lb]{\large{{$s_2$}}}
    \Text(141,44)[lb]{\large{{$p_4$}}}
    \Text(157,39)[lb]{\large{{$s_3$}}}
    \Text(182,46)[lb]{\large{{$p_5$}}}
    \Line(61,-18)(61,6)
    \SetWidth{0.5}
    \Vertex(98,8){2.83}
    \Photon(98,9)(64,8){2}{6}
    \SetWidth{1.0}
    \Line(61,35)(61,10)
    \SetWidth{0.5}
    \Vertex(61,9){2.83}
    \SetWidth{1.0}
    \Line(179,-23)(179,1)
    \Text(191,4)[lb]{\large{{$\gamma(t_3)$}}}
    \Text(73,-6)[lb]{\large{{$t_1$}}}
    \Text(93,-10)[lb]{\large{{$G_2$}}}
    \Text(117,-7)[lb]{\large{{$t_2$}}}
    \SetWidth{0.5}
    \Photon(102,9)(139,9){2}{6}
    \Vertex(143,9){2.83}
    \Photon(145,8)(182,8){2}{6}
    \Vertex(180,6){2.83}
    \SetWidth{1.0}
    \Line(143,11)(143,35)
    \Text(139,-12)[lb]{\large{{$G_2$}}}
    \Text(163,-7)[lb]{\large{{$t_3$}}}
    \Line(180,10)(180,34)
    \Text(60,-33)[lb]{\large{{$p_1$}}}
    \Text(179,-34)[lb]{\large{{$p_6$}}}
    \Text(97,44)[lb]{\large{{$p_3$}}}
    \Text(59,45)[lb]{\large{{$p_2$}}}
    \Text(27,8)[lb]{\large{{$\gamma(t_1)$}}}
  \end{picture}
\caption{Linear triple-Regge limit with with internal vertices  
$G_2(t_1,\kappa_1,t_2)$ and  $G_2(t_2,\kappa_2,t_3)$, which are determined by factorization from the 5-point amplitude.}
\label{fig:lineartripleregge6}
\end{center}
\end{figure}
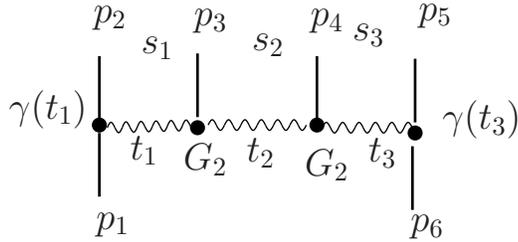

Let us next examine   the multi-Regge limit, Fig. \ref{fig:lineartripleregge6},
\bea
&&s_1, s_2, s_3 \rightarrow - \infty;\;\;\; \nonumber\\
&&t_1,t_2,t_3,
\kappa_1, \kappa_2 <0 \;\;\;\;{\rm fixed}
\eea
 One again finds $s/s_2\rightarrow \infty$,   and 
\be
u_1\rightarrow 0\;,\quad u_2\rightarrow  0\;,\quad u_3\rightarrow 1
\ee
In this limit, the dilogarithms again do not contribute, and 
we obtain the desired multi-Regge limit, 
 \be\label{eq:multiregge6}
 {\cal A}_6 \sim 
\left(\frac{s_1}{t_1}\right)^{\alpha(t_1)}\left(\frac{s_2}{t_2}\right)^{\alpha(t_2)}
\left(\frac{s_3}
{t_3}\right)^{\alpha(t_3)}\,  
 \ee
as in Eq. (\ref{eq:multiregge6-2}) and also depicted in Fig. \ref{fig:lineartripleregge6}.

It is worth commenting that, in arriving at the multi-Regge limit, (\ref{eq:multiregge6}), cancellation of 
quadratic terms in $\ log s_j$ must occur. We note that, the net contribution to $\log M_6$  from 
$A_{div}$ and $L_{6}$ is
\be
 - \frac{f(\lambda)}{4 }  \ln (-s_1 ) \ln (-s_3) \;.
 \ee
 This crossed-term is cancelled when $Q_6$ is taken into account.

\begin{figure}[th]
\begin{center}
\includegraphics[width = 0.70\textwidth]{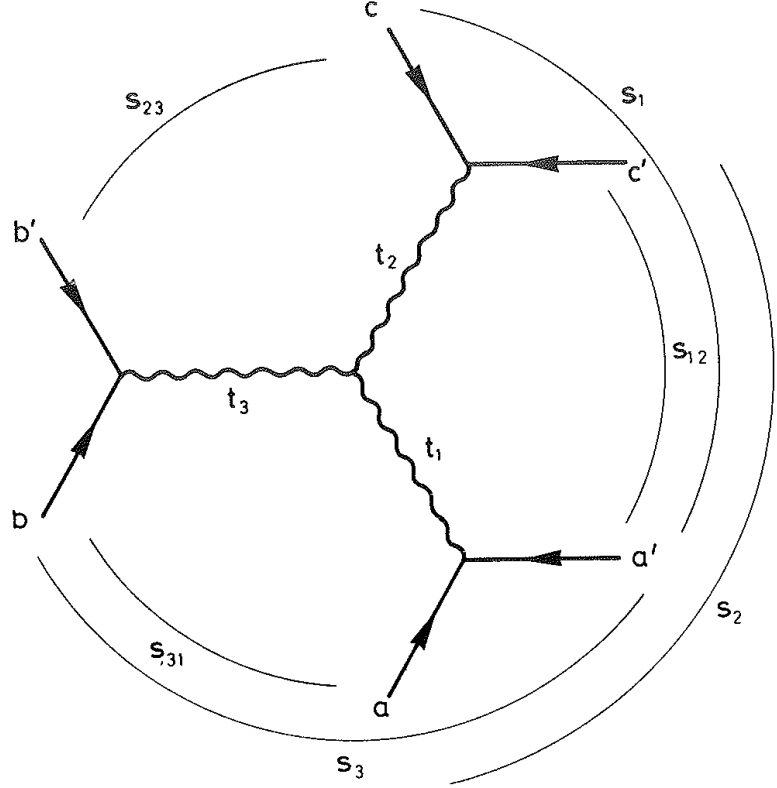}
\end{center}
\caption{Triple-Regge limit of six-point amplitude.}
\label{fig:triple}
\end{figure}

We now turn to a new class of  ``poly-Regge limits". We consider the kinematical variables described 
in 
Fig.~\ref{fig:triple}, which for $n=6$, is referred to as 
``triple-Regge" limit ~\cite{Detar:1971gn,Detar:1972nd}. Note that this is not the "linear multi-Regge limit" discussed above and more 
generally  in Sec. \ref{sec:regge}, but rather a symmetric  limit involving a set of 
three pairs of near-collinear momenta~\footnote{ The ``triple-Regge" limit is conventionally 
associated with single-particle inclusive production, e.g., high-mass diffractive dissociation. The limit 
discussed here is related but yet different from this more conventional usage. To reach the usual 
inclusive limit, another so-called ``helicity pole" limit\cite{Detar:1971gn,Detar:1971dj,Detar:1972nd} is required.}. 

The variables of Fig.~\ref{fig:triple} can be expressed in term of BDS variables
as follows, (identified in counter clockwise order), 
\[     t^{[2]}_{a'} = s_{12}  ; \; \;t^{[2]}_c = s_{23} ; \; \;t^{[2]}_b = s_{31} ;
 \; \; t^{[2]}_{a} = t_1 ; \; \;t^{[2]}_{c'} = t_2 ; \; \;t^{[2]}_{b'} = t_3 ;
 \; \; t^{[3]}_{a'} = s_1 ; \; \;t^{[3]}_{a} = s_{2} ; \; \;t^{[3]}_b = s_{3} 
\]
However only 8 of the 9 variables are independent invariants. Define
\be
\eta_{ij} = \frac{s_{ij}}{s_i s_j}\;.
\ee
A constraint can be written involving $\eta$'s and $t$'s, and it can be used to 
eliminate one of the $\eta$'s ~ \cite{Detar:1972nd}.
Proceeding as in Sec.~\ref{sec:four} and Sec.~\ref{sec:five}, 
we find the exact result
\bea \label{eq:6BDS}
{\cal A}_6 &=& \mbox{const} \; A_{tree} \prod^3_{i=1} A^2_{div}(t_i)
\E{\frac{f(\lambda)}{16}[2Q_{6} +  \lambda(\log s_{12}, \log s_{23}, \log s_{31})]} \nn
&\times&\E{[\frac{f^{(-1)}(\lambda)}{8 \epsilon}  + \frac{g(\lambda)}{4} ][ \log( s_{12}/
t_1) + \mbox{cyclic}] +
\frac{f(\lambda)}{16} [ \log^2(t_1/\mu^2) + \mbox{cyclic}]} \nn
&\times&\E{-\frac{f(\lambda)}{8} [\log(\eta_{13}/\eta_{23}) \log( s_{13}/s_{23}) + 
\mbox{cyclic}] +\frac{f(\lambda)}{16} [\log^2(\eta_{13}/\eta_{23})  + \mbox{cyclic}]}\nn
&\times&\E{- \frac{f(\lambda)}{8} [ {\rm Li}_2(1-t_1 \eta_{23})+{\rm cyclic}]}  
\eea
where $\lambda(a,b,c) = a^2 + b^2 + c^2 - 2 ab -2 bc - 2 ac$.  Note that the cross ratios, which enter 
through the  arguments of the dilogarithms, again remain fixed and finite. One can show again that 
the $\log^2s_{ij}$ type terms  in (\ref{eq:6BDS})  cancel.    The triple-Regge limit most appropriate to Fig~\ref{fig:triple} is
\bea
&&s_1 , \;\;s_2 , \;\;s_3 , \;\;s_{12} , \;\;s_{23} , \;\; s_{31}  \; 
- \rightarrow \infty \nn
&&t_1 , \;\;t_2 , \;\;t_3 , \;\;\eta_{12} , \;\;\eta_{23} , \;\; \eta_{31} < 0  \; ,
\quad \mbox{fixed}
\label{3regge}
\eea
i.e., $u_1,u_2,u_3$ are again fixed. One finds that  (\ref{eq:6BDS}) can again be expressed 
simply in terms of the gluon Regge trajectory, 
with the desired triple Regge behavior~\cite{Detar:1972nd},
\be \label{eq:6regge}
{\cal A}_6 = \prod^3_{i=1} A^2_{div}(t_i) \prod^3_{j=1} \left(\frac{-s_j}
{\mu^2}\right)^{\alpha(t_j)}
\; F(t_i, \eta_{ij})
\ee

Finally, we note that Regge behavior for 6-point gluon amplitude has also been addressed in 
\cite{McGreevy:2007kt} from a different perspective.

\section{The general BDS $n$-gluon amplitude}
\label{sec:general}

We next  demonstrate that the findings for $n=6$ in the single-Regge limits and the maximal linear 
multi-Regge limit can be generalized directly to $n>6$. For $n>6$, type-I single-Regge refer to 
either $s_1$ becoming large or  the equivalent limit where $s_{n-3}$ becomes large, 
whereas type-II will refer to the limit where one of the remaining $s_k$'s becoming large.

The BDS formula for general $n$ is  ${\cal A}_n={\cal A}_{tree,n}M_n$, 
\bea
&&\ln M_n= \sum_{i=1}^n \ln A_{div}(t_i^{[2]})+{\cal F}_n^{(1)}(0)\;,
\label{md}
\eea
where  $A_{div}$ is given by (\ref{eq:div}) and 
${\cal F}_n^{(1)}(0)=(f(\lambda)/8)(Q_n+D_n+L_n +  (3n/2) \zeta_2 )$ 
is obtained from the finite part of the 1-loop  ordered n-point amplitude, with 
\bea
Q_n &=& -\sum_{i=1}^{n} \sum_{r=2}^{[n/2]-1} \ln\left(- \frac{t^{[r]}_i}{- t^{[r+1]}_{i}}\right)
\ln\left( \frac{-t^{[r]}_{i+1}}{-t^{[r+1]}_{i}}\right)\label{eq;Qn}
\eea
and, depending on $n$ even or odd,  
\bea
D_{2m}&=&
-\sum_{i=1}^{2m}\sum_{r=2}^{m-2}{\rm Li}_2\left(1-\frac{t^{[r]}_{i} t^{[r+2]}_{i-1}}{ 
t^{[r+1]}_i t^{[r+1]}_{i-1}}\right)
- \sum_{i=1}^{m}{\rm Li}_2\left(1-\frac{t^{[m-1]}_{i} t^{[m+1]}_{i-1}}{ 
t^{[m]}_i t^{[m]}_{i-1}}\right)   \label{eq:deven2}\\
L_{2m}&=& - \frac{1}{4}\sum_{i=1}^{2m}
\ln\left(\frac{-t^{[m]}_{i}}{ -t^{[m]}_{i+m+1}}\right)\ln\left(\frac{-t^{[m]}_{i+1}}{ -t^{[m]}_{i+m}}\right) =
\frac{1}{2}\sum_{i=1}^{m}
\ln^2\left(\frac{-t^{[m]}_{i}}{ -t^{[m]}_{i+1}}\right)\\
\label{even}
 D_{2m+1}&=&
-\sum_{i=1}^{2m+1}\sum_{r=2}^{m-1}{\rm Li}_2\left(1-\frac{t^{[r]}_{i} t^{[r+2]}_{i-1}}{ 
t^{[r+1]}_i t^{[r+1]}_{i-1}}\right)\label{eq:dodd2}\\
L_{2m+1}&=& -\frac{1}{2}\sum_{i=1}^{2m+1}
\ln\left(\frac{-t^{[m]}_{i}}{ -t^{[m]}_{i+m+1}}\right)
\ln\left(\frac{-t^{[m]}_{i+1}}{ -t^{[m]}_{i+m}}\right)
\label{odd}
\eea
As in the case of $n=6$, we first note the appearance of dilogarithm, with arguments depending on 
the cross-ratios
\be
B_{i,r} = \frac{t^{[r]}_{i} t^{[r+2]}_{i-1}}{ 
t^{[r+1]}_i t^{[r+1]}_{i-1}}=\frac{x_{i,j}^2x_{i-1,j+1}^2}{x_{i,j+1}^2x_{i-1,j}^2} =u(i,j;i-1,j+1)\;,
\label{crossrati}
\ee
where $j=i+r$. 
Note that $B_{i,2}$ is 
a generalization for $n\geq 6$ of the cross-ratios $u_1,u_2,u_3$, defined for $n=6$.
We will see later another kind of generalized cross-ratio,
$u_{k,n}$ ($n=2m$ or $2m+1$), appearing as a constraint in the 
Regge limits. In the next section cross-ratios will appear in a different 
context, giving a possible parametrization of the n-point amplitudes. 
Like for  the case of $n=6$, we can show that all these cross-ratios remain bounded in various Regge limits of interests. 

In what follows, we will  focus  on the even amplitudes, $M_{2m}$, $3\leq m$, and, the discussion for 
$n$ odd, $n=2m+1$, will be also be done in Appendix B. Except for lengthy algebra, nothing note-
worthy is found for the odd amplitudes.

We will first analyze the maximal {\bf linear multi-Regge limit}, where all $s_i$, $i=1, 2, \cdots, $ 
$n-3$ are large, and find that we obtain the correct  Regge behaviour.

Using  the fact that,  in the multi-Regge limit,  the BDS invariant $t_i^{[r]}$,  for $2\leq i< i+r\leq n-1$, takes on the factorized form, Eq. (\ref{bdsmr}), we show in Appendix B that  these cross-ratios approach either $0$ or $1$, just as the case for $n=6$, discussed in Sec. \ref{sec:six}. For example, we have, in the multi-Regge limit,
\bea
B_{1,m-1}&=&\frac{t^{[m-1]}_{1} t^{[m+1]}_{2m}}{ 
t^{[m]}_1 t^{[m]}_{2m}} =  \left( \frac{ t_{m-2} }{t_{m-1}}\right)\left( \frac{ t^{[m-1]}_{m+1}}{ 
 t^{[m]}_{m}}  \right)  \simeq  \left( \frac{ t_{m-2} }{t_{m-1}}\right)\left( \frac{ 1}{ 
\kappa_{m-1}s_{m-1} } \right)
\sim\frac{1}{s_{m-1}}\rightarrow 0\nn
B_{i,r}& =& \frac{t^{[r]}_{i} t^{[r+2]}_{i-1}}{ 
t^{[r+1]}_i t^{[r+1]}_{i-1}} = \left(\frac{t^{[r]}_{i} }{ t^{[r+1]}_{i-1}}\right) \left(  \frac{ t^{[r+2]}_{i-1}}
{ t^{[r+1]}_i }\right) \simeq \left(\frac {\kappa_{i-1}s_{i-1}}{1}\right) \left(\frac{ 1 }{\kappa_{i-1} s_{i-1}} 
\right) 
\rightarrow 1\; ,\nn
&&\quad\quad\hskip 2.5in  \;\; 3\leq i < i+r  \leq 2m-1\;.
\eea
It follows that these dilogarithms will not affect the leading Regge behaviour. A similar statement 
applies for 
the single Regge limits, except there the $B_{i,r}$'s 
cover what is defined in Appendix A as type 1, 2 and 3 cross-ratios
and thus can take on finite values that differ from 0 and 1.

A necessary condition for Regge behaviours is the cancellation  of all quadratic terms, both direct, 
$\ln^2 s_j$, and crossed, $\ln s_i\ln s_j$, $i\neq j$, $i,j =1, 2, \cdots n-3$, for $\log M_n$.

Let us  first express the contribution from $A_{div}$ as
\bea
\sum_{i=1}^{2m}\ln A_{div}(t_i^{[2]})&=& -\frac{f(\lambda)}{16}
\left[\sum_{i=1}^{2m-3}\ln^2 s_i+\ln^2 s\right]   +O(1) \nn
&=&
-\frac{f(\lambda)}{16}\left[\sum_{i=1}^{2m-3}\ln^2 s_i+\ln^2(s_1s_2....s_{2m-3})\right]+O(\ln s_i)
\eea
where we have used that, in the multi-Regge limit, the dependent variable $s$ is 
\be
s\simeq b_0\;  s_1s_2....s_{2m-3} 
\ee
where $b_0= \Pi_{i=1}^{2m-4} \kappa_i$.

We next turn to the contributions from  $L_{2m}$. To find them, we need to  know how  ratios 
$t_{i}^{[m]}/t_{i+1}^{[m]}  $, $1 \leq i \leq m$,  behave in the multi-Regge limit.  
Recall, from (\ref{bdsmr}), that $t^{[m]}_i$ is  linear in $s_j$ if $j$ falls in the range $[i-1, m+i-3]$.   It 
follows that
\be
t^{[m]}_{i} \simeq \beta_{i-1}\; s_{i-1}s_{i} \cdots s_{m+i-4}s_{m+i-3}\;, \label{eq:tmrlimit}
\ee
where $\beta_i= \Pi_{j=i }^{m+i-3} \kappa_j$, and
\be
\frac{t^{[m]}_{i}}{t^{[m]}_{i+1} }\sim   \frac{ s_{i-1}}{s_{m+i-2}}  
\ee
Since Eq. (\ref{eq:tmrlimit}) is valid for $2\leq i\leq 2m-r$ and $2\leq r\leq m$, it leads  to
\bea
 L_{2m}
 &\simeq &\frac{1}{2} \left[    \ln^2(\frac{t_{m-1}}{s_1...s_{m-1}}) +\sum_{i=1}^{m-2} \ln^2\frac{ s_i }
{ s_{m+i-1}}+\ln^2(\frac{s_{m-1}s_{m}...s_{2m-3}}{t_{m-1}}) \right] + O(\ln s_i)\nn
&=&\frac{1}{2}\left[ \sum_{i=1}^{2m-3}\ln^2 s_i+\ln^2(s_1s_2....s_{2m-3}) +({\rm crossed \;\;terms})
\right] + O(\ln s_i)
\eea
where we have made use of  the fact that $t_1^{[m]}=t^{[m]}_{m+1}=t_{m-1}$.   Upon  adding 
$(f(\lambda)/8)L_{2m}$ and  $\sum_{i=1}^{2m}\ln A_{div}(t_i^{[2]})$,  all quadratic terms of the form 
$\ln^2 s_j$ cancel,  leaving only  crossed terms as uncanceled contributions to $\ln M_{2m}$, 
\bea
\ln M_{2m} &= &- \frac{f(\lambda)}{8}\left[  \sum_{i=1}^{m-2}\ln s_i\ln s_{m+i-1} 
+\ln (s_1...s_{m-2})\ln (s_m...s_{2m-3})\right]   \nn
&+& \frac{f(\lambda)}{8} Q_{2m} + O(\ln s_i)\;.\label{quadraticAL}
\eea
For $m=3$, we have
\be
\ln M_{6} = - \frac{f(\lambda)}{4 }  \ln (s_1) \ln( s_3)   + \frac{f(\lambda)}{8} Q_{6} + O(\ln( s_i))\; , 
\label{lnsilnsj}
\ee
which agrees with what we found earlier.

For the contribution from $Q_{2m}$ we need to know how $\ln \frac{t^{[r]}_{i}}{t^{[r+1]}_{i} }\ln 
\frac{t^{[r]}_{i+1}}{t^{[r+1]}_{i} }$ behaves in the multi-Regge limit. 
That is, we need to extend  (\ref{eq:tmrlimit}) to $ t^{[r]}_{i}$ for $ 1\leq i\leq  2m-r-1$ and 
$2\leq r\leq m$, and next to a wider range, i.e., $1\leq i\leq 2m+1$ and $2\leq r\leq m-1$.  This is done in Appendix B. 
>From 
(\ref{tlimit}), we find
\bea
Q_{2m} &\simeq & -
\sum_{r=2}^{m-1}\left[ \sum_{i=1}^{2m-r-2} \ln s_i\ln s_{r+i-1} -\ln{s_1...s_{2m-r-2}}\ln s_{2m-r-1}\right] 
\nn
&-&  \sum_{r=2}^{m-1}\left[ \sum_{i=1}^{r-2}\ln s_i \ln s_{2m-r+i-1} - \ln s_{r-1}\ln {s_r...s_{2m-3}}\right]
+O(\ln s_j)
\eea
Note that this is a sum of quadratic products,  $\ln s_i\ln s_j$, $i\neq j$. For instance, for $m=3$,
\be
Q_{6} \simeq 2\ln s_1 \ln s_3 + O(\ln s_j) \;,
\ee
which cancels the corresponding term in (\ref{lnsilnsj}). For general $n=2m$, after expanding as a 
sum over  $\ln s_i \ln s_j$, and after a bit of algebra, 
these quadratic terms from $Q_{2m}$ cancel exactly against those obtained above, 
(\ref{quadraticAL}). 

We are now in the position to verify Regge behaviour by finding terms linear in $\ln s_j$, thus 
verifying that the same 
Regge trajectory function emerges, as dictated by Regge factorization. In order to carry out this 
analysis, we must 
keep track of the coefficients of proportionality in (\ref{s-multiregge}),  (\ref{eq:tmrlimit}), etc. We have 
done this
explicitly in the multi-Regge limit, but the algebra is a bit involved, so we will show some of it in the 
Appendix.
We will instead
show this analysis for the single-Regge limits, while leaving out some of the algebraic details. 
Note that although the algebra is quite different from the multi-Regge case, we in fact obtain the 
same result 
for the leading $s_k$ behaviour in both cases. If all the single-Regge limits give the expected 
trajectories, 
the multi-Regge behaviour will also give the correct result.

We thus finally obtain that
\be
\ln M_{2m} \simeq \sum_{k=1}^{2m-3} (\alpha(t_k)-1) \ln(- s_k ) + O(1)
\ee
A similar analysis can also be carried out for the 
$O(1)$ terms. It follows that $M_{2m}$ can now be put into the expected multi-Regge power law 
behaviour, 
(\ref{eq:multireggegeneral}).
Details of this analysis, together with that for $n=2m+1$, are shown in Appendix B.

We now address the {\bf single Regge limits}. In the case of $n=6$, we have distinguished two types 
of single-Regge limits.  For type-I, one of the two vertices is elastic, involving only two particles. 
We will provide a unified treatment here where an arbitrary  $s_k$ goes to infinity, with other 
$s_r$'s fixed. Because of symmetry, we can restrict to $1\leq k\leq m-1$, and type-I Regge 
corresponds to $k=1$.

To be precise, the single-Regge limit now  corresponds to 
\be
s_k \rightarrow  - \infty, 
\ee
with  $t_i$, $\kappa_j$,  and all other $s_l$ negative and  fixed. As discussed  in Sec. \ref{sec:regge} and in 
Appendix A, those BDS invariants, $t_i^{[r]}$, which ``cross" the Regge line, labelled by $t_k$, (see 
Figs. \ref{fig:limits} and \ref{fig:cross_ratio}), will also go to infinity. That is,  $t_i^{[r]}\rightarrow \infty$, 
with $ \frac {t_i^{[r]}}{s_k}$ fixed, 
\bea
&&{\rm for} \;\; 2\leq i\leq k+1\;\;, \quad   \quad {\rm if} \quad k+3\leq i+ r \leq 2m\nn
&&{\rm for } \;\; k+2\leq i\leq 2m\;\; ,   \quad   \quad  {\rm if} \quad 2\leq i+r-2m \leq k +1\;.
\eea
In particular, the dependent variable $s$ also goes to infinity, with $s/s_k$ fixed.  

With these preliminaries, we now repeat what we have done earlier in the cancellation of  the 
quadratic term, $\ln^2 s_k$,  while keeping track of terms linear in $\ln s_k$. From $L_{2m}$ and 
$A_{div}$,  the $\ln^2 s_k$ term indeed cancels, and  we obtain, (after a bit of algebra),
the contribution to $\ln M_{2m}$ which is linear in $\ln s_k$, 
\be
 \left[\frac{f(\lambda)}{8} \ln \left(\frac{ t_2^{[m]}t^{[m]}_{k+1}\mu^4}{t_{m-1}s s_k t_{k+2}^{[m]}}\right)  
  +\frac{g(\lambda)}{2}+\frac{f^{-1}(\lambda)}{4\epsilon}\right] \ln s_k
\ee

The analysis of the terms coming from $Q_{2m}$ is understandably quite involved, and we will not 
reproduce it here.  As expected, it does not contain a $\ln s_k^2$ term, and the result is a contribution   
$O(\ln s_k)$
\be
\frac{f}{8}\ln \left[\left( \frac{\mu^2}{-t_k^2}\right)
\left(\frac{t_{m-1}s_kst_{k+2}^{[m]}}{ t_2^{[m]}\mu^4t_{k+1}^{[m]}}\right)
\left(\frac{s_k s}{t_2^{[k+1]} t_{2m}^{[k+1]}}\right) \right]\ln s_k
\ee
Adding up these two contributions, we obtain, after re-expressing $s_k=t_{k+2}^{[2]}$ and 
$s=t_{2}^{[2m-2]}$
\be
\ln M_{2m} \simeq (\alpha(t_k)-1)  \ln s_k +\frac{f}{8}\; \ln \left(\frac{ t_{k+1}^{[2]}  \; t_{2}^{[2m-2]}}
{t_2^{[k+1]}t_{2m}^{[k+1]}}\right)\ln s_k
\ee

For the type-I single-Regge limit, with $k=1$ (or $k=2m-3$), 
the last factor cancels identically, and, as in the case of $n=6$, 
we obtain the desired Regge trajectory.

Next consider type-II single-Regge. Note that the combination in the last term is a cross-ratio, 
$2\leq k \leq 2m-4$, 
(a different generalization of $u_3$ to the case of $n\geq6$ than $B_{i,r}$)
\be
u_{k,2m}=\frac{ t_{2}^{[2m-2]}\; t_{k+1}^{[2]}  }{t_2^{[k+1]}t_{2m}^{[k+1]}}= u(2, 2m; k+1, k+3)\;.
\ee
For $m=3$ and $k=2$, one easily verify that $u_{2,6}=u_3$ introduced in (\ref{crossr}). 
Since this cross-ratio approaches 1  in the single-Regge limit (see eq. (\ref{uconstra})), 
\be
u(2, 2m; k+1, k+3) \rightarrow 1
 \ee
it follows that 
\be
{\cal A}_{2m} \sim \left(\frac{s_k}{t_k}\right)^{\alpha(t_k)}\label{singleregge}
\ee
for both type-I and type-II   single Regge limits, as advertised. The same cross-ratio constraint, 
$u_{k,2m}\rightarrow1$ appears in the Regge trajectory for the multi-Regge limit.

\section{Regge behaviour and dual conformal symmetry}
\label{sec:conformal}

In this section we describe the interplay between Regge behaviour of the n-gluon 
amplitude and the solution of the conformal 
Ward identities for the n-cusp Wilson loop \cite{Drummond:2007au}. Drummond et al. 
\cite{Drummond:2007aua} and 
Brandhuber et al. \cite{Brandhuber:2007yx}
proposed a duality between the n-sided
polygon Wilson loops and the n-point gluon amplitudes, in part from a perturbative 
analysis for n=4,5 \cite{Drummond:2007aua} and general n \cite{Brandhuber:2007yx}.
At 1 loop, 
the duality is found to be exact. Further, the Alday and Maldacena proposal \cite{Alday:2007hr} for 
the strong coupling gluon 
amplitude implies      the validity of the  duality as a consequence of the trivial geometric equality of 
the AdS space (where one calculates the minimal surface of the Wilson loop) and of the T-dual AdS 
space (where one calculates the minimal surface giving the gluon amplitude). However, to define the 
Wilson loop or its dual one needs a regulator, which makes the duality less trivial.  This duality is also 
broken at finite temperature \cite{Ito:2007zy}.

Drummond et al. \cite{Drummond:2007cf,Drummond:2007au} further proved a 
  conformal Ward identity for the polygon  Wilson loops for all 
orders in perturbation theory. Recently Komargodski 
\cite{Komargodski:2008wa} proved the anomalous Ward identity of Drummond et al. for the strong-
coupling (AM) dual of the gluon amplitude. One writes the expectation value of the Wilson loop 
 $W(C_n)$ as 
\be
\ln W_n (C_n)=Z_n +F_n^{(WL)}
\ee
which separates the divergent factor $Z_n$ from the finite part of the Wilson loop, 
and $C_n$ is the 
same contour that appears in (\ref{eq:npoint}). They  propose that 
\be
\sum_{i=1}^n (2 x_i^{\mu} x_i\cdot \partial_i -x_i^2 \partial_i^{\mu})F_n^{(WL)}=
\frac{1}{2}\Gamma_{cusp}(\lambda)
\sum_{i=1}^n\ln \left(\frac{x_{i,i+2}^2}{x_{i-1,i+1}^2}\right)x_{i,i+1}^{\mu}
\label{ward}
\ee
where
\be
x^2_{i,i+r}=(k_i+...+k_{i+r-1})^2;\;\;\;\;
k_i=x_i-x_{i+1}\label{identi}
\ee
thus $x_{i,i+r}^2$ is $t_i^{[r]}$ in the case of the dual gluon amplitude, and $
\Gamma_{cusp}=Cf(\lambda)$ is the
cusp anomalous dimension that appears in the previous sections. The most 
general solution to (\ref{ward}) for n=4
and n=5 is 
\be
F_4^{(WL)}=F_4^{(BDS)}+{\rm constant};\;\;\;\;\;\;
F_5^{(WL)}=F_5^{(BDS)}+{\rm constant}
\label{wlbds}
\ee
where 
\be
\exp F_n^{(BDS)}=[A_n^{(BDS)}]_{finite}
\ee
Equation (\ref{wlbds}) lends support to the BDS conjecture for n=4 and 5, and led to 
the proposed duality 
between the n-cornered Wilson loop and the n-gluon amplitude.

The general solution of (\ref{ward}) for n=6 is
\be
F_6^{(WL)}=F_6^{(BDS)}+f(u_1,u_2,u_3)\label{solu}
\ee
where $f(u_1,u_2,u_3)$ is an arbitrary function of the three cross-ratios 
\be
u_1=\frac{x_{13}^2 x_{46}^2}{x_{14}^2x_{36}^2};\;\;\;
u_2=\frac{x_{24}^2x_{15}^2}{x_{25}^2x_{14}^2};\;\;\;
u_3=\frac{x_{35}^2 x_{26}^2}{x_{36}^2x_{25}^2}\label{crossratios}
\ee
It is important to note that the dilog terms $D_{6,i}$ in (\ref{eq:d6}) are functions of the 
cross-ratios only, 
as we saw already in (\ref{crossr}),
and as such do not contribute to the right-hand side of (\ref{ward}). Recently, 
Drummond et al. \cite{Drummond:2007bm} found that in general
$f(u_1,u_2,u_3)\neq $ constant, from a two-loop calculation of $F_6^{(WL)}$.

Now consider (\ref{ward}) from the point of view of this paper, where we claim that the Regge and multi-Regge behaviour of the  n-gluon amplitudes is a necessary property of the theory. Note that 
the dilog terms, $D_6$, in 
the last line of (\ref{eq:6BDS}) are {\em finite} and nonzero in the various Euclidean Regge limits considered in Sec.
\ref{sec:six}. This observation then 
generalizes to the fact that in 
the single-Regge limits and triple-Regge limits, it is plausible that
\be
f(u_1,u_2,u_3)={\rm finite}\label{funct}
\ee
in the Euclidean region for {\em any} function of the cross-ratios, as the cross-ratios are finite in these limits.
Taking (\ref{solu}) together with (\ref{funct}),  this implies that $F^{WL}_6$ has the same Regge and  
behaviour as  $F_6^{(BDS)}$, where $F_6^{WL}$ is the solution of  (\ref{ward}).

The line of argument in the preceding paragraphs generalizes to $n>6$. Using the results of \cite{Drummond:2007au},
we obtain that the solution to (7.2) for $n\geq 7$ is 
\be
F_n^{(WL)}= F_n^{(BDS)}+f(u_1,u_2,...,u_p)\label{resolution}
\ee
where $f(u_1,...,u_p)$ is an arbitrary function of the distinct cross-ratios. It is shown in Sec. 6 that the 
Regge limits of  $F_n^{(BDS)}$ have the expected Regge behavior in the Euclidean region. For example, the function of 
cross ratios  $B_{i,r}$, (\ref{crossrati}), (generalizations of $u_1,u_2,u_3$ for $n\geq 7$) remain finite 
in the limit leading to (\ref{singleregge}). We therefore conclude that the Euclidean Regge limits considered in 
this paper do not distinguish between $F_n^{WL}$, $F_n^{BDS}$, and $F_n$, where the latter is the 
actual n-gluon amplitudes. Further, if the functions of the cross-ratios in (\ref{solu}) and 
(\ref{resolution}) remain finite in the Regge limits, then $F_n^{(WL)}$ is also not distinguished from 
$F_n^{BDS}$ by Regge behavior in the Euclidean region.

\section{Concluding remarks}
\label{sec:concl}

It is expected that ${\cal N}=4$ SYM theory exhibits stringy properties, albeit
with infinite string tension as required by the conformal symmetry of the
theory. This view is based on the 't Hooft $1/N$ expansion, as well as the  AdS/CFT
correspondence. Indeed Alday and Maldacena  have used the
AdS/CFT correspondence for the Wilson loop in strong coupling to predict
the color-ordered n-gluon scattering amplitude for large N, $SU(N)$ ${\cal N} = 4$
SYM theory. For the 4-point function, the AM construction is in agreement with  the
BDS conjecture for the 4-gluon MHV planar scattering amplitude. Further it
has been shown that the $n=4$ amplitude can be cast exactly as a Regge 
amplitude with a 
large N Regge trajectory function valid to all orders in perturbation theory, 
presented in (\ref{eq:alpha}).

In this paper we have examined the Regge and multi-Regge behavior of the BDS 
conjecture for $n \ge 5$ in  the Euclidean region.  It was found that the BDS conjecture is consistent with the 
Regge limit taken in this region. A crucial tool in this conclusion is an understanding of the behavior of various cross-
ratios in Regge limits.

It is known from the recent work of Drummond et al.\cite{Drummond:2007bm} that the hexagonal 
Wilson loop 
differs from the BDS 
conjecture for the 6-gluon amplitude by a non-constant function of the three cross-
ratios. Given the 
basic assumption of this paper, that the n-gluon amplitudes should exhibit Regge 
and multi-Regge behaviour, we 
showed in section 7 that if the function in (\ref{solu}) and (\ref{resolution}) remain finite in the Regge 
limits, then  Regge behavior in the Euclidean region does not distinguish between $F_n^{WL}$, $F_n^{(BDS)}$, or $F_n$ 
where $F_n$ is the n-gluon amplitude. Continuation to the physical region enables one to examine this issue further \cite{Bartels:2008ce,Brower:2008ia}. 

We should emphasize that there is considerably more that can be done
using Regge limits to test and constrain conjectures for the n-point
${\cal N} =4$ SYM amplitudes. We have not exhausted the full repertoire of
limits determined by the Regge hypothesis or considered the analytic continuation to the physical region, or the full
constraints of factorization and analyticity on residue functions. For example, we have not discussed  the constraints of overlapping singularities and  the issue of analyticity in $\kappa_i$, an important but subtle issue~\cite{Brower:1974yv,Detar:1972nd,Detar:1971dj}, as well as the crossing relations in other related multi-Regge limits~\cite{Detar:1971dj} which can play a significant role in the BFKL program~\cite{BL,Kuraev:1977fs} for the high energy limit of ${\cal N}=4$ SYM in the vacuum  channel with cylinder topology~\cite{Brower:2006ea,Brower:2007qh,Brower:2007xg}. Also
for $n\geq 6$, we have not investigated contributions below the leading
trajectory, which in the planar limit are expected to be simple poles in the $J$-plane,
free of Regge cuts.

Finally, given  the crucial role played by the various cross-ratios and their limits in our analysis, one 
may conjecture that the limits of cross-ratios are a central issue to be addressed in achieving Regge 
behavior  in any conformal theory, and perhaps other theories as well.

{\bf Note added:}~\footnote{We are grateful to the referee for the detailed report on our paper, which  led to this Note added.} 

Since the appearence of this paper (BNST-I) on the archives, there have been several papers   extending the present   investigation~\cite{Bartels:2008ce,Brower:2008ia,DelDuca:2008jg,Bartels:2008sc} in interesting   directions.  While they generally go beyond the scope of this article, they do contribute valuable   additional information of relevance to the underlying question of Regge properties for BDS   amplitudes. In particular, these papers have considered the important issue of the multi-Regge behavior continued into the physical region.  In Ref. \cite{Bartels:2008ce}, which was posted shortly after our paper appeared, Bartels, Lipatov and Sabio Vera (BLSV-I)   studied  the linear multi-Regge behavior for six-gluon amplitudes in the physical region. For certain color configurations, non-factorization was found.  The issue of analytic continuation back to the physical scattering region for all relevant color configurations is also the focus of our subsequent paper~\cite{Brower:2008ia} on Regge limits of BDS (BNST-II). We compare and contrast features of the BDS amplitudes with that from flat-space string theory, paying particular attention to proper handling of threshold singularities of multi-Regge amplitudes, thus maintaining causality relations in the course of continuation back to the physical region.

The continuation discussed by BLSV-I involves expanding in the dimensional regulator $\epsilon\rightarrow 0$ and then analytically continuing the finite part of the 1-loop amplitude (that appears in the exponent of the BDS ansatz) according to a causal prescription.  The BLSV-I continuation among other things leads to a subtle diverging term $\sim 2\pi i \ln(1-u_3)$, where $u_3$ is a conformally invariant cross-ratio approaching 1 in the Regge limit, as well as an extra term, which is not a function of conformally invariant cross-ratios.  As it stands, the BLSV-I continuation does not permit ``naive factorization'' in the form of Eq. (5.17) of our paper. There are many subtleties, some of which we discuss in \cite{Brower:2008ia}. For example, we note that, for flat-space string theory, physical region factorization in the multi-Regge limit applies only to amplitude of definite ``signature'', not to specific subset  of planar amplitudes identified in BLSV-I as non-factorizable. Further analysis was carried out in   BLSV-II \cite{Bartels:2008sc}.  We also note that a paper by Del Duca, Duhr and Glover \cite{DelDuca:2008jg} appeared after \cite{Brower:2008ia}.   They have  considered the possibility of analytical   continuation with $\epsilon \neq 0$ fixed, taking the Regge limit before the   $\epsilon\rightarrow 0$ limit, which leads to a result differing from that of BLSV-I.  In their approach,  the anomalous term of BLSV-I does not appear.   Further discussion of these issues will be addressed in an updated version of BNST-II.

Finally the results of Bern et al. \cite{Bern:2008ap} and Drummond et al. \cite{Drummond:2008aq}   are also  interesting in the present context. They show numerically that the BDS ansatz for the six-gluon amplitude fails for finite (non-limiting) kinematics.   On the other hand, the duality between MHV amplitudes and  Wilson-loops  have been shown to hold  to 2 loops and 6 external legs, (for the parity even part in \cite{Bern:2008ap,Drummond:2008aq}, for the parity odd part in \cite{Cachazo:2008hp}.)  In view of the theorems (both all-loop perturbative \cite{Drummond:2007au} and strong coupling \cite{Komargodski:2008wa}) proving dual conformal symmetry of the Wilson-loop, the deviation from the 6-gluon BDS amplitudes that was found in  \cite{Bern:2008ap,Drummond:2008aq} { must} be a function of cross-ratios. 
It is also worth noting  that Refs. \cite{Berkovits:2008ic,Beisert:2008iq} proved that dual conformal symmetry arises from a combination of bosonic plus fermionic T dualities, and \cite{Drummond:2008vq} embedded dual conformal symmetry in dual superconformal symmetry.

\newpage

{\bf Acknowledgements:} The work of RCB was supported by the Department of 
Energy under
Contract.~No.~DE-FG02-91ER40676. HN would like to thank Katsushi Ito and 
Alexei Morozov for discussions. HN's research  has been done with partial support 
from  MEXT's program "Promotion of Environmental Improvement for 
Independence of Young Researchers" under the Special Coordination Funds for 
Promoting Science and Technology. HJS wishes to thank Lance Dixon for 
conversations and correspondence. He also wishes to acknowledge the  stimulating atmosphere of 
the 
2007 Brown string meeting, which played an important role in forming this 
collaboration. HJS's research is supported in part by the DOE under grant DE-
FG02-92ER40706.  CIT would like to thank E. M. Levin for discussion and prior 
collaboration and also thank G. Korchemski for correspondence. His research is supported in part by 
the U.~S.~Department of 
Energy under  Contract DE-FG02-91ER40688, TASK A. In particular, we would like to thank Marcus 
Spradlin and Anastasia Volovich for pointing out a crucial error in the original version of this 
manuscript.

\newpage

\appendix

\section{BDS variables and constraints}

\label{app:BDSvar} 

In this Appendix we describe the variables used by BDS and constraints between 
them, in general and in the Regge limits, as well as a parametrization 
for them used in the multi-Regge limit in the text.

The variables used by BDS are $t_i^{[r]}\equiv (k_i+...+k_{i+r-1})^2$
or $x_{i,i+r}^2$ in the Wilson loop dual notation of
\cite{Drummond:2007au}. (All indices are defined mod n.)  There are
$n(n-3)/2$ such cyclic invariants, since momentum energy conservation implies
$t_i^{[r]} = t^{[n-r]}_{i+r}$.  However, the number of independent Lorentz
invariant parameters in the amplitude should be $3n-10$ (a simple way
to see this is that there are 2 parameters for the 4-point amplitude,
$s$ and $t$, and each new on-shell momentum adds another 3).  That
means that only the 4-point and 5-point amplitudes (when
$n(n-3)/2=3n-10$) are described by independent $t_i^{[r]}$'s (2 and 5
of them, respectively), for $n>5$ there are constraints between them.

To understand these constraints, we first look at the first nontrivial
case, the 6-point amplitude. Then there are 9 $t_i^{[r]}$ variables,
but there should be 8 parameters, so there is one constraint between
them.  The variables can be chosen to be $t_i^{[2]}, i=1,...,6$ and
$t_1^{[3]}, t_2^{[3]}, t_3^{[3]}$.

Now let us describe the constraint.\footnote{We thank Lance Dixon for 
communication and  for providing us with  his program for the constraint equation.}
Since we have a 6-point amplitude, there are 6 momenta $k_i$, but momentum 
conservation fixes one of them, 
e.g. $k_6=-(k_1+...+k_5)$, so there are 5 momenta that can be 
independently varied. But invariance under the Poincare group restricts further the 
number of parameters appearing 
in the amplitude. Since we are in a 4 dimensional space, the 5 momenta must be 
linearly dependent. 
Therefore $\exists \alpha_i$ constants such that $\sum_i \alpha_i k_i=0$ (4 
equations for the 4 components). Multiplying
with $k_j$ we get 5 dependent equations, therefore
\be
P_5\equiv \det M_{ij}=0; \;\;\; M_{ij}\equiv (k_i\cdot k_j),\;\; i,j=1,..,5
\ee
This is a constraint equation in terms of $k_i\cdot k_j$, but it turns out that all of 
them can be expressed in terms
of $t_i^{[r]}$'s. Since $k_i^2=0$, we have
\be
t_i^{[r]}=\sum_{i\leq l,m\leq i+r-1} k_l\cdot k_m\label{tofk}
\ee
and in particular $t_i^{[2]}=2k_i\cdot k_{i+1}$ for $i=1,..,4$. In addition, we have
\bea
 2k_1\cdot k_3=t_1^{[3]}-t_1^{[2]}-t_2^{[2]}; &&
 2k_2\cdot k_4=t_2^{[3]}-t_2^{[2]}-t_3^{[2]}\nonumber\\
 2k_3\cdot k_5=t_3^{[3]}-t_3^{[2]}-t_4^{[2]}; &&
 2k_5\cdot k_1= t_2^{[3]}-t_5^{[2]}-t_6^{[2]}\nonumber\\
 2k_1\cdot k_4=t_2^{[2]}+t_5^{[2]}-t_1^{[3]}-t_2^{[3]}; &&
2k_2\cdot k_5=t_3^{[2]}+t_6^{[2]}-t_2^{[3]}-t_3^{[3]}
\eea
These complete the 10 nontrivial elements of $M_{ij}$, thus $P_5$ is expressed as 
a homogeneous polynomial of degree 
5 in the $t_i^{[r]}$ variables. In principle it could be an identity (i.e., $P_5\equiv 0$) 
and not a constraint, 
but expanding it explicitly shows it is indeed a constraint. We will not reproduce it 
here, since it will fill 
half a page and we would learn nothing new. 

We now turn to the general case of an $n>5$ point amplitude. We need $(n-4)(n-5)/
2=\begin{pmatrix} n-4&\\2&\\ 
\end{pmatrix}$ constraints, obtained in a similar way. Among the $n$ momenta  
pick a basis for 4 dimensional space, say $k_1,k_2,k_3,k_4$.
>From the remaining $n-4$ momenta, pick  one to be expressed from the others by 
momentum conservation, 
e.g. $k_n=-(k_1+...+k_{n-1})$, and call it $k_i$. Then pick another one, $k_j$, who 
will be expressed as a linear 
combination of the basis $k_1,..,k_4$. As above, we obtain the constraint $\det 
M_{i''j''}=0$, $M_{i''j''}=k_{i''}\cdot 
k_{j''}$, where $i''\in (1,2,3,4,j)$.

Now $2k_{i'}\cdot k_{j'}$, where $i'=1,...,n$ except $k_i$,
 can be expressed as a linear combination of the $t_i^{[r]}$'s as we did for the 6-
point amplitude, i.e.
\be
2k_{i'}\cdot k_{j'}
=\sum_{r,k}\alpha_{r,k}t_k^{[r]}\label{soft}
\ee
This can be seen as follows. There are $n(n-1)/2$ nontrivial 
values for $k_l\cdot k_m$, l,m=1,...n, since $k_i^2=0$ (thus 
$l=m$ is excluded). 
By picking out 
$k_i$ and writing it as minus the sum of the other momenta (by momentum 
conservation), we can express $n-1$ of the 
$k_l\cdot k_m$ variables
as a function of the others, and we are left with $(n-1)(n-2)/2=n(n-3)/2+1$ 
independent ones. But we still have 
one constraint left
\be
\sum_{i',j'=1, i',j'\neq i}^{n}k_{i'}\cdot k_{j'}=(\sum_{i'=1,i'\neq i}^nk_{i'})^2=(-k_i)^2=0
\ee
therefore we can eliminate another one of the $k_{i'}\cdot k_{j'}$ as a function of the 
others, and we are left with 
$n(n-3)/2$, exactly the number of variables $t_i^{[r]}$. That means that we can now 
invert the relation (\ref{tofk}) 
and obtain (\ref{soft}).

Therefore, the polynomial constraint $\det M_{i''j''}=0$ is again expressed in terms 
of the $t_i^{[r]}$'s. But we 
needed to pick two momenta, $i$ and $j$, out of the $n-4$ dependent ones, 
therefore we can do this in 
$(n-4)(n-5)/2$ ways, obtaining the needed $(n-4)(n-5)/2$ constraints (actually, 
choosing first $i$ and then $j$ 
or vice versa seems to be different, but by considering only one case we already 
obtain all the needed constraints, 
therefore the reversed case will give the same). 

In conclusion, the variables $t_i^{[r]}$ are restricted by $(n-4)(n-5)/2$ complicated 
5-th order polynomial 
constraints. This will be cumbersome for the treatment of limits of $n$-point 
amplitudes. Luckily, in the Regge 
limits, the constraints greatly simplify, and one can deal with them systematically. 

As described in the text, in the linear multi-Regge limit one can choose a set of 
$3n-10$ independent parameters 
among the $t_i^{[r]}$'s that describe the physics well, specifically
\be
s_i=t_{i+1}^{[2]}, \; i=1,...,n-3\;\;\;\hspace{1cm} t_r=t_1^{[r+1]}=t_{r+2}^{[n-r-1]},\; 
r=1,...,n-3
\ee
and 
\be
\Sigma_i=t_{i+1}^{[3]},\; i=1,...,n-4
\ee
that describe the limit as in Fig.\ref{fig:limits}. The advantage is that they can be varied 
independently, and the 
behaviour of the rest of the $t_i^{[r]}$'s can be found in terms of them.
Some 
constraints become easy to write down
in the Regge limits. 

We now show a systematic way to find such {\bf constraints in the single 
Regge limit}, for $s_k\rightarrow \infty$. They are related with the definition of the generalized 
cross ratios, whose usefulness we have found throughout this paper.

The Regge line for $s_k$ separates the momenta into right movers $k_i^+\rightarrow \infty$
(for momenta on the left of the line) and left movers $p_j^-\rightarrow \infty$ (for 
momenta on the right of the line). Cross ratios correspond to two momentum invariants divided 
by other two momentum invariants, and can be described as falling into 3 classes, in the 
$s_k$ single Regge limit:

\begin{itemize}
\item 1. All 4 momentum invariants involve either only right ($k$'s) or only left ($p$'s) momenta.
\item 2. In two of the 4 invariants we have both left and right momenta.
\item 3. All 4 momentum invariants involve both left and right momenta.
\end{itemize}

Type-1 cross ratios are finite variables, and
obviously appear either in the left or the right factorized 
function in the Regge limit.

Type-2 cross ratios also become finite variables of only the right-moving or left-moving  momenta, 
thus appear in 
either the left or right factorized function in the Regge limit. For concreteness, we will 
look at the case where the variables are on the left. Define
\be
u(i,j;p,q)=\frac{(k_i+...+k_{j-1})^2(k_p+...+k_{q-1})^2}{(k_i+...+k_{q-1})^2(k_p+...+k_{j-1})^2}=
\frac{x_{i,j}^2x_{p,q}^2}{x_{i,q}^2x_{p,j}^2}
\ee
where $2\leq i<p<q-1\leq k+1< j-1\leq n-1$. Also define $P^{\mu}=\sum_{r=i}^{k+1}k_r^{\mu}$,
$Q^{\mu}=\sum_{r=k+2}^{j-1}k_r^{\mu}$ and $p^{\mu}=\sum_{r=p}^{k+1}k_r^{\mu}$. Then in 
the limit,
\be
u(i,j;p,q)\simeq\frac{(k_p+...k_{q-1})^2(P^+Q^-)}{(k_i+...k_{q-1})^2(p^+Q^-)}
=\frac{(k_p+...k_{q-1})^2P^+}{(k_i+...k_{q-1})^2p^+}
\ee
which is a finite function of only right-moving variables, as stated.

Type-3 cross ratios are the important ones, that will give constraints. We define similar
variables
\be
u(i,j;p,q)=\frac{(k_i+...+k_{j-1})^2(k_p+...+k_{q-1})^2}{(k_i+...+k_{q-1})^2(k_p+...+k_{j-1})^2}=
\frac{x_{i,j}^2x_{p,q}^2}{x_{i,q}^2x_{p,j}^2}
\ee
but where now $2\leq i<p<q-1<j-1\leq n-1$, and $p\leq k+1,k+2\leq q-1 $, and $P^{\mu}=
\sum_{r=i}^{k+1}
k_r^{\mu}$, $Q^{\mu}=\sum_{r=k+2}^{j-1} k_r^{\mu}$, $p^{\mu}=\sum_{r=p}^{k+1}k_r^{\mu}$, 
$q^{\mu}=\sum_{r=k+2}^{q-1} k_r^{\mu}$,  we get in the limit
\be
u(i,j;p,q)\simeq \frac{(P^+Q^-)(p^+q^-)}{(P^+q^-)(p^+Q^-)}=1\label{uconstra}
\ee
which therefore represents a constraint, that has been used in the text.

Further kinematic simplifications can be achieved in the {\bf linear multi-Regge limit}, where, after 
supplying a minus sign for each out-going longitudinal momenta,
\be
-k_2^+>> -k_3^+ >> \cdots - k_i^+>>\cdots>>-k_{n-2}^+ >> -k_{n-1}^+
\ee
In this limit,  for $2\leq i<j\leq n$,
\be
x_{i,j}^2= t_i^{[j-i]}=(\sum_{r=i}^{j-1} k_r)^2 \simeq   2(\sum_rk_r)^+ (\sum_rk_{r})^- \simeq 2 k^+_i k^-
_{j-1} +O(1) \;.\label{eq:strongordering}
\ee
(When $i=1$, $x_{1,j}^2 =t_j$, which is fixed in this limit, and above approximation is no longer 
valid.)   It  follows that we can derive a recursive relation for invariants $t_i^{[r]}=[i,i+r]$, 
\be
t_{i+1}^{[j-i+2]} \simeq  s_{i} \kappa_{i} \;  t_{i+2}^{[j-i+1] }  \;,\quad 
t_{i+1}^{[j-i+2]} \simeq      t_{i+1}^{[j-i+1]} \;   s_j \kappa_{j-1}  \;, 
\ee
for $1 \leq i<j\leq n-3$, 
and  from which one has
\be
[i,j]= t_{i} ^{[j-i]} = \kappa_{i-1} \kappa_{i}\cdots  \kappa_{j-4}s_{i-1}s_{i}\cdots s_{j-4} 
s_{j-3}\label{bdsmr}
\ee
This  also means that, in this limit,
\be
s   \simeq  \kappa_1 \kappa_2\cdots
\kappa_{n-4} {s_1s_2\cdots
  s_{n-3}}
\label{eq:s-mr}
\ee
as mentioned in Sec. \ref{sec:regge}, Eq. (\ref{s-multiregge}).

\section{Regge Limits for $n>6$: details}
\label{app:appendixB}

Here we provide details which have been left out of Sec. \ref{sec:general}. We begin 
with the treatment of the {\bf linear multi-Regge limit}.  We first discuss how
kinematic constraints in the multi-Regge limit  can be used to study  the
dilogarithms terms in (\ref{eq:deven2}) and (\ref{eq:dodd2}), which depend 
on the generalized cross-ratios $B_{i,r}$. We show specifically
that $B_{i,r}$ remain bounded in the linear 
multi-Regge limit. We have
\be
B_{i,r}=u(i,i+r; i-1,i+r+1)=\frac{x_{i,i+r}^2x_{i-1,i+r+1}^2}{x_{i,i+r+1}^2x_{i-1,i+r}^2}=\frac{t^{[r]}_{i} t^{[r
+2]}_{i-1}}{ 
t^{[r+1]}_i t^{[r+1]}_{i-1}}
\ee
where $D_n$  for n even and odd can be expressed as 
\bea
D_{2m}&=&-\sum_{i=1}^{2m} \sum_{r=2}^{m-2}{\rm Li}_2\left(1-B_{i,r}\right)-
\sum_{i=1}^{m} {\rm 
Li}_2\left(1-B_{i,m-1}\right)
\label{eq:deven}\\
 D_{2m+1}&=&
-\sum_{i=1}^{2m+1}\sum_{r=2}^{m-1}{\rm 
Li}_2\left(1-B_{i,r}\right)\label{eq:dodd}
\eea
Note that $B_{i,2}$  becomes $u_1,u_2,u_3$ (\ref{crossr}) for $n=6$, thus $B_{r,i}$ are 
generalizations for $n\geq 6$ and $1\leq i\leq n$ of these variables, analyzed in Sec. \ref{sec:regge}.

We will first analyze the {\bf n=2m} case.  
The behaviour for some of the cross ratios can be found easily, e.g., using (\ref{bdsmr}). More 
directly, using (\ref{eq:strongordering}), we get
\bea
B_{i,j-i}
&= &\frac{x_{i,j}^2x_{i-1,j+1}^2}{x_{i,j+1}^2x_{i-1,j}^2}\simeq   \frac{(k_i^+ k_{j-1}^-)(  k_{i-1}^+ 
k_{j}^-)}{(k_i^+k_{j}^-)(k_{i-1}^+ k_{j-1}^- )}=1\;, \label{eq:Blimits}
\eea
provided that  $i, i-1, j , j-1\neq 1$.   Thus all cross-ratios in (\ref{eq:deven}) and (\ref{eq:dodd}), 
approach
\be
B_{i,r}\rightarrow 1\;,
\ee
with the exception of those  involving  $x_1$. (When  the initial momenta $k_1$ and $k_{n}$ enter, 
the approximation, Eq. (\ref{eq:strongordering}) is no longer valid.)  For
$B_{i,m-1}$, $1\leq i\leq m$,  the allowed special cases are $i=1$ and $i=2$, giving
$u(1, r+1; 2m, r+2)$ and $  u(2,r+2 ; 1 , r+3)$
respectively.  Like the cross-ratios $u_1$ and $u_2$, they vanish in the multi-Regge limit. Explicitly, 
we find from (\ref{bdsmr}),
\be
B_{1,m-1}= 
\left( \frac{ t_{m-2} }{t_{m-1}}\right)\left( \frac{ t^{[m-1]}_{m+1}}{ 
 t^{[m]}_{m}}  \right)  
\sim\frac{1}{s_{m-1}}\rightarrow 0;\;\;\;\;
B_{2,m-1}=
 \left( \frac{ t_{m} }{t_{m-1}}\right)\left( \frac{ t^{[m-1]}_{2}}{ 
 t^{[m]}_2}  \right) 
\sim\frac{1}{s_{m-1}}\rightarrow 0
\ee
For $B_{i,r}$,  one of $i, i-1, i+r , i+r-1$ being $ 1$, gives 
$i=1,2,2m-r, 2m-r+1$, with $2\leq r\leq m-2$. We find, again using  (\ref{bdsmr}),
\bea
&&B_{1,r}=
\left( \frac{ t_{r-1} }{t_{r}}\right)\left( \frac{  t^{[r+2]}_{2m}}{ 
 t^{[r+1]}_{2m}}  \right)  \sim\frac{1}{s_r}\rightarrow 0,\;\;\;\;
B_{2,r}=
\left( \frac{ t_{r+1} }{t_{r}}\right)\left( \frac{ t^{[r]}_{2}}{ 
 t^{[r+1]}_{2}}  \right) 
\sim  \frac{1}{s_r}
\rightarrow 0,\nonumber\\
&&B_{2m-r,r}=\left( \frac{ t_{2m-r-3} }{t_{2m-r-2}}\right)\left( \frac{t^{[r]}_{2m-r}}{ 
t^{[r+1]}_{2m-r-1} }  \right)  \sim \frac{1}{s_{2m-r-2} } 
\rightarrow 0,\nonumber\\
&&B_{2m-r+1, r}=
\left( \frac{ t_{2m-r-1} }{t_{2m-r-2}}\right)\left( \frac{t^{[r+2]}_{2m-r}}{ 
t^{[r+1]}_{2m-r}}  \right) \sim \frac{1}{s_{2m-r-2}}\rightarrow 0\;.
\eea

We therefore have shown that  $B_{m-1,i}$ and $B_{r,i}$ are bounded in the multi-Regge limit
 (they approach either 1 or 0). 
 Therefore $D_{2m}$ remains finite and the dilog's in (\ref{eq:deven}) will not affect the leading 
behavior for $\ln M_{2m}$.

As described in section \ref{sec:regge} and 
the Appendix A, it is convenient to specify the BDS variables  by 
$t_i^{[r]}$ for $r<m$,  $i=1,...,2m$,   and $t^{[m]}_i$ for $i=1,...,m$. However, there are 
constraints among them. 
A set of  $3n-10$ independent invariants can 
be taken to be 
$s_i=t_{i+1}^{[2]}$  for $i=1,...,2m-3$, $t_r=t_1^{[r+1]}=t_{r+2}^{[n-r-1]}$ for 
$r=1,...,2m-3$, and 
$\Sigma_i=t_{i+1}^{[3]}$ for $i=1,...,2m-4$. (Equivalently, instead of $\Sigma_i$, we 
can use $\kappa_i=\Sigma_i/s_i s_{i+1}$.)  
This set is most convenient for describing the maximal linear multi-Regge limit, where  $s_i$ and $
\Sigma_i$ are going 
to infinity, with $t_i$ and $\kappa_j$ fixed.   
>From  the constraint  (\ref{eq:s-mr}), $s$, as a dependent variable, also goes to infinity, with the 
following ratio fixed
\be
b_0=  \frac{s}{s_1...s_{2m-3}}
\simeq \kappa_1\kappa_2\cdots\kappa_{2m-4}  \;.
\ee

In order to understand how contributions from $L_{2m}$ and $Q_{2m}$  behave in various Regge 
limits,   we need to know the behaviour of all $t_i^{[r]}$'s. First, $t_1^{[r]}=t_{r+1}^{[2m-r]}=t_{r-1}$ is 
fixed.
Then there are two  special cases for $r$. The first group, for $r=2$  
$t_i^{[2]}$, $i=1,...,2m$, is familiar, given by 
$\{t_1,s_1,...,s_{2m-3},t_{2m-3},s\}$. The second group,  $\{t_i^{[m]}\}$,  $i=1,...,m$, will be denoted 
for convenience as 
$\{t_{m-1},\gamma_1,...,\gamma_{m-1}\}$, i.e.,
\be
\gamma_i = t_{i-1}^{[m]}, \quad i=2, 3,\cdots, m-1
\ee
 As described in section 3, since each invariant $\gamma_i$ crosses   dotted 
lines $m-1$ times, they go to infinity 
as 
\be
\gamma_i \sim s_{i-1} s_i \cdots s_{i+m-2}
\ee
By a similar analysis for the general case of $2<r<m$, 
the following set of ratios remain finite, 
\bea
b_0&=&\frac{s}{s_1...s_{2m-3}};\;\;\;\;
\beta_i^{[r]}=\frac{t_i^{[r]}}{s_{i-1}...s_{r+i-3}}\;\;\;\; , \quad 2\leq i \leq 2m-r \;,\nn  
\beta_{2m-r+i}^{[r]}&=&\frac{t_{2m-r+i}^{[r]}}{s_{i-1}...s_{2m-r+i-3}}\;\;\;\; ,  \quad 2\leq i\leq r;\;\;\;\;\;
\beta_i=\frac{\gamma_i}{s_i...s_{i+m-2}}\label{tlimit}
\eea
where $\beta_{i+1}^{[m]}=\beta_{i+m+1}^{[m]}=\beta_i$, $\beta_{2m}^{[2]}=b_0$ and $
\beta_i^{[2]}=1$
for i=2 to $2m-2$.

Let us next turn to $L_{2m}$,  $A_{div}$, and $Q_{2m}$.  We have shown in Sec. \ref{sec:general} 
that
 in the multi-Regge limit all quadratic terms in $\ln s_j$, both direct, $\ln s_j^2$, and crossed, $\ln s_i 
\ln s_j$, $i\neq j$, cancel, when contributions from $A_{div}$, $L_{2m}$, and $Q_{2m}$ are 
combined. We therefore need  to focus on the terms linear in $\ln s_j$.

The $\ln s_j$ terms from  $L_{2m}$ are
\bea
&&\frac{f}{8}\left[
\ln (s_1... s_{m-1})\ln \frac{\beta_1}{t_{m-1}}+\sum_{i=1}^{m-2}\ln\frac{s_i}{s_{m+i-1}}\ln\frac{\beta_i}
{\beta_{i+1}}+ \ln(s_{m-1}...s_{2m-3})\ln\frac{\beta
_{m-1}}{t_{m-1}}\right]+O(1)\nn
\eea
where we have not bothered to keep the additional contributions,
\be
\sum_{k=1}^{2m-3}\left[\frac{g(\lambda)}{2}+\frac{f^{-1}(\lambda)}{4\epsilon}\right] \ln s_k\;.
\ee
The $\ln s_i$ terms from $\sum A_{div}$,  are 
\be
-\frac{f}{8}\ln (s_1...s_{2m-3})\ln b_0 +0(1).\label{gterms}
\ee
with $b_0 = s/(s_1...s_{2m-3})$ fixed in all the Regge limits.
Adding them up and expressing in terms of $\ln s_i$ separately, we get
\bea
&&\hspace{-1.2cm}\frac{f}{8}\left[\sum_{k=1}^{m-2} \ln s_k\ln\left(\frac{\beta_1}{b_0t_{m-1}}
\frac{\beta_k}{\beta_{k+1}}\right)+\ln s_{m-1}\ln \left(\frac{\beta_1\beta_{m-1}}{b_0t^2_{m-1}}\right)+  
\sum_{k=1}^{m-2} \ln s_{m+k-1} \ln \left(\frac{\beta_{m-1}}{b_0t_{m-1}}
\frac{\beta_{k+1}}{\beta_k}\right)\right]\nonumber\\
\label{lnsterms}
\eea
Now we turn to the  terms linear in $\ln s_j$ coming from $Q_{2m}$. The analysis is tedious, 
so we will only reproduce the results.
For the terms proportional to $\ln s_1$ (relevant for the  type-I single Regge limit), we obtain
\bea
&&
-\frac{f}{8}\ln s_1\ln \left[\frac{(t_1)^2}{t_{m-1}}\frac{\beta_2^{[m]}\beta_{m+2}^{[m]}}
{\beta_{2m}^{[2]}\beta_{m+3}^{[m]}}\right]
\eea
Adding this to the corresponding $\ln s_1$ term in (\ref{lnsterms}) and (\ref{gterms}), we find an exact 
cancellation, except for the factor
\be
\left[-\frac{f}{4}\ln t_1+\frac{g(\lambda)}{2}+\frac{f^{-1}(\lambda)}{4\epsilon}\right]\ln s_1 \;,
\ee
which is the required term to give the correct Regge trajectory.

We now turn to the general term, $\ln s_k$, with $k\neq 1, 2m-3$. The contribution from $Q_{2m}$ is 
found to be 
\be
-\frac{f}{8}\ln s_k\ln \left[ \frac{(t_k)^2}{t_{m-1}}\frac{\beta_k\beta_1}{\beta_{k+1}b_0}
\frac{\beta_2^{[k+1]} \beta_{2m}^{[k+1]}}{b_0}\right]
\ee
Adding this to the $\ln s_k$ terms in (\ref{lnsterms}), after re-expressing 
$\beta's$ in terms of BDS invariants, we get 
\be
(\alpha(t_k)-1) \ln s_k    + \frac{f}{8}\ln s_k \ln u_{k,2m} + O(1) \label{calcu}
\ee
where
\be
u_{k,2m}=  \frac{b_0}{\beta_2^{[k+1]}\beta_{2m}^{[k+1]}}=
\left[  \frac{t_{2}^{[2m-2]}t_{k+1}^{[2]}} {t_2^{[k+1]}t_{2m}^{[k+1]}}\right]=
\frac{x_{2,2m}^2x_{k+1,k+3}^2}{x_{2,k+3}^2x_{2m,k+1}^2}=u(2,2m;k+1,k+3)\;.
\ee
The single-Regge limit constraint (\ref{uconstra}) is also valid in the multi-Regge limit, implying  
\be
u_{k,2m}=1
\ee
$M_{2m}$ can now be put into the expected multi-Regge power law behaviour, 
(\ref{eq:multireggegeneral}).

We see that as promised in section 6, we have obtained the same  $\ln s_k$ 
terms (including the extra term involving the 
constraint $u_{k,2m}$) from the multi-Regge limit as we did from the single-Regge limit, even though 
the algebra 
looks different (different contributions add up to the same result).

We now turn to the case with {\bf n=2m+1}, and make a similar analysis. We will focus on the 
{linear multi-Regge limit}. As for the n=2m case, we find that the single Regge limit can be 
deduced from the multi-Regge limit (even though the calculation is quite different).
Then $t_i^{[2]}$ are given by $t_1,s_1,...,s_{2m-2},s$ for $i=1,...,2m+1$, $t_1^{[r]}=t_{r+1}^{[2m+1-
r]}=t_{r-1}$ 
is fixed and the 
behaviour of the rest of $t_i^{[r]}$ for
$2<r\leq m$ is defined by the fact that the following quantities are finite
\bea
&& \frac{s}{s_1...s_{2m-2}}=b_0=\beta_{2m+1}^{[2]};\;\;\;\;\beta_i^{[r]}=\frac{t_i^{[r]}}{s_i...s_{r+i-3}},
\;\;\; 2\leq i\leq 2m+1-r\nonumber\\
&& \beta_{2m+1-r+i^{[r]}}=\frac{t_{2m+1-r+i}^{[r]}}{s_{i-1}...s_{2m-r+i-2}},\;\;\; 2\leq i\leq r;
\;\;\;\;\; \beta_i^{[m]}\equiv \beta_i\;.
\eea
We again evaluate the quantities $B_{i,r}$ in (\ref{odd}) and find that they are 0 or 1 in 
the multi-Regge limit: $B_{1,r}=B_{2,r}=B_{2m+1-r,r}=B_{2m+2-r,r}=0$, while the rest of $B_{i,r}$
are 1, by the kinematical constraint $u(i,i+r;i-1,i+r+1)=1$.
Therefore, it follows that $D_{2m+1}$ remains finite in the linear multi-Regge 
limit. 

Let us turn  next to the quadratic terms $\ln s_j\ln s_j$. From $L_{2m+1}$, one has 
\bea
L_{2m+1} &\rightarrow &
\frac{1}{2}\left[\sum_{i=1}^{2m-2} \ln^2 s_i -\sum_{i=1}^{m-2}\ln s_i \ln(s_{m+i-1} s_{m+i}) - 
\sum_{i=1}^{m-2}\ln s_{m+i}\ln (s_is_{i+1}) \right ]  \nn
&+&\frac{1}{2}\left[ \ln^2(s_1\cdots s_{m-1})+\ln^2 (s_m\cdots s_{2m-2})-\ln s_1\ln s_m-\ln s_{m-1}\ln 
s_{2m-2} \right]
\nn
&+& \left[  \ln s_m \ln (s_1\cdots s_{m-1})   +\ln s_{m-1}\ln (s_m\cdots s_{2m-2})\right]  
+O(\ln s_i)
\eea
The divergent terms, $A_{div}$,  give
\bea
\sum_{i=1}^{2m}\ln A_{div}(t_i^{[2]})&\rightarrow& -\frac{f(\lambda)}{16}[
\sum_{i=1}^{2m-2}\ln^2 s_i +\ln^2(s_1\cdots s_{m-1})+\ln^2 (s_m\cdots s_{2m-2})\nonumber\\
&&+2\ln (s_1\cdots s_{m-1})\ln (s_m\cdots s_{2m-2})]\;,
\eea
thus again there are uncancelled $\ln s_i \ln s_j$ terms remaining in $\ln M_{2m
+1}$. Adding these two contributions, one is left with
\bea
&&\hspace{-1cm}\frac{f}{8}\left[\ln s_{m-1}\ln (s_m\cdots s_{2m-3})-\sum_{i=1}^{m-2} \ln s_i \ln(s_{m
+i-1} s_{m+i}) - \ln (s_1\cdots s_{m-1})\ln (s_{m+1}\cdots s_{2m-2})\right]\nn
\eea
Just like the case for $n$ even, these quadratic terms cancel against those from $Q_{2m+1}$. 

We now turn to the $O(\ln s_i)$ terms.
After a long calculation, we find that the $O(\ln s_i)$ terms coming from $L_{2m+1}$ and $\ln A_{div}
$ can be written as 
\bea
&-&\frac{f}{8}  \sum_{k=1}^{m-2} \left[ \ln s_k\ln \left(t_{m-1}\frac{b_0\beta_{m+3+k}}{\beta_{k
+1}\beta_{m+3}}\right)       +\ln s_{m+k}
\ln \left(t_m\frac{b_0\beta_{m+k+2}}{\beta_{k+3}\beta_{2m+1}}\right)\right]   \nn
&-&\frac{f}{8}\left[\ln s_{m-1}\ln \left((t_{m-1})^2\frac{b_0}{\beta_m\beta_{m+1}}\right)
 +\ln s_m \ln \left((t_m)^2\frac{b_0}{\beta_{2m+1}\beta_3}\right)  \right]\nn
&+& \sum_{k=1}^{2m-2}\left[\frac{g(\lambda)}{2}+\frac{f^{-1}(\lambda)}{4\epsilon}\right] \ln s_k\;.
\eea

The terms coming from $Q_{2m+1}$ are even more involved. The $\ln s_1$ terms, relevant for 
the type I single Regge limit, are found to be 
\be
-\frac{f}{8}\ln s_1 \ln \left(\frac{(t_1)^2}{t_{m-1}}\frac{\beta_2\beta_{m+3}}{b_0\beta_{m+4}}\right)
\ee
Adding them to the previous $\ln s_1$ terms, we get the required term that gives the correct 
Regge trajectory, 
\be
(\alpha(t_1)-1) \ln s_1 +O(1)
\ee
and as before, we did not need to use the constraints to get this result.

The generic $\ln s_k$ term from $Q_{2m+1}$ with $k\leq m-2$ is found to be 
\be
-\frac{f}{8}\ln s_k\ln \left(\frac{(t_k)^2}{t_{m-1}}\frac{\beta_2^{[k+1]}\beta_{2m+1}^{[k+1]}}{b_0}
\frac{\beta_{k+1}\beta_{m+3}}{b_0\beta_{m+k+3}}\right)
\ee

Adding to the $\ln s_k$ terms from $L_{2m+1}$ and divergent terms we obtain 
\be
(\alpha(t_k)-1) \ln s_k-\frac{f}{8}\ln s_k\ln \left(\frac{\beta_2^{[k+1]}\beta_{2m+1}^{[k+1]}}{b_0}\right)
=(\alpha(t_k)-1) \ln s_k+\frac{f}{8}\ln s_k\ln u_{k, 2m+1}
\ee
where 
\be
u_{k,2m+1}= \frac{b_0}{\beta_2^{[k+1]}\beta_{2m+1}^{[k+1]}}=
 \left[  \frac{t_{2}^{[2m-1]}t_{k+1}^{[2]}} {t_2^{[k+1]}t_{2m+1}^{[k+1]}}\right]=
 \frac{x_{2,2m+1}^2 x_{k+3,k+1}^2}{x_{2,k+3}^2x_{2m+1,k+1}^2}=u(2,2m+1;k+1,k+3)
\ee
and again, in the linear multi-Regge limit, the single-Regge kinematic constraint (\ref{uconstra}) still 
holds, and
implies that 
\be
u_{k,2m+1}\rightarrow 1
\ee
so the extra term cancels, and we obtain the correct Regge trajectory. The case $k=m-1$ is 
found to be included in the general case (even though the terms coming from $Q_{2m+1}$
look different), and the terms with $k\geq m$ work by symmetry.

In the single Regge limit we obtain the same corresponding $s_k$ terms as from the multi-Regge 
limit 
(just as in the $n=2m$ case). Again type I is satisfied without use of the constraints, but for the 
type II single Regge limit ($k\neq 1,2m-2$) we have the extra term 
\be
 \frac{f}{8}\ln u_{k,2m+1}
\ee
that cancels due to the use of the constraints (\ref{uconstra}), with 
$2\leq i<p<q-1<j-1\leq 2m-1$ and $p\leq k+1,q-1\geq k+2$, applied for $i=2,p=k+1,q=k+2,j=2m$.

For completeness, we write down explicitly the definition of the single Regge limit. As usual, 
we have, for a given $k\leq m-1$
\be
s_k\rightarrow \infty;\;\;\;\; s\rightarrow \infty;\;\;\;\;
\frac{s_k}{s}=\tilde{b}={\rm fixed}
\ee
but we also have other $t_i^{[r]}$ going to infinity, with 
\be
\frac{t_i^{[r]}}{s}=k_i^{[r]}={\rm fixed}
\ee
which occurs if 
\bea
&& k+1\geq i \geq 2 \;\;\;\;{\rm and}\;\;\;\; 2m+1\geq i+r\geq k+3\;\;\;\;\;{\rm or}\nonumber\\
&& 2m\geq i \geq k+2\;\;\;\;\; {\rm and}\;\;\;\;\ 2m+k+2\geq +r\geq 2m+3
\eea

\bibliographystyle{utphys}
\bibliography{bds_v3}

\end{document}